\UseRawInputEncoding
\documentclass[lettersize,journal]{IEEEtran}
\usepackage{array}
\usepackage[caption=false,font=normalsize,labelfont=sf,textfont=sf]{subfig}
\usepackage{cite}
\usepackage{amsmath,amssymb,amsfonts}
\usepackage{algorithmic}
\usepackage{graphicx,color}
\usepackage{textcomp}
\usepackage{amsmath,amsfonts}
\usepackage{paralist}
\usepackage{algorithm}
\usepackage{algorithmic}
\usepackage{array}
\usepackage[caption=false,font=normalsize,labelfont=sf,textfont=sf]{subfig}
\usepackage{textcomp}
\usepackage{stfloats}
\usepackage{url}
\usepackage{verbatim}
\usepackage{graphicx}
\usepackage{cite}
\usepackage{xcolor}
\usepackage{svg}
\usepackage{pdfpages}
\usepackage{diagbox}
\def\BibTeX{{\rm B\kern-.05em{\sc i\kern-.025em b}\kern-.08em
    T\kern-.1667em\lower.7ex\hbox{E}\kern-.125emX}}

\newtheorem{theorem}{Theorem}
\newtheorem{lemma}{Lemma}

\begin{document}

\title{Deep Learning Aided Beamforming for Downlink Non-Orthogonal Multiple Access Systems}

\author{Georgios Konstantopoulos, Graduate Student Member, IEEE,\\ and Yves Lou\"et \\ CentraleSupélec, Campus Rennes, 35510,
 Cesson Sevigne, France
\thanks{This paper was submitted and is under review by IEEE Open Journal of the Communications Society}
\thanks{This work was carried out within the framework of the French collaborative project "Covera5Ge" supported by DGA
and whose partners are CentraleSupélec, ENENSYS
Technologies, and Siradel.}

}

\markboth{Journal of \LaTeX\ Class Files,~Vol.~X, No.~X, August~20XX}%
{Shell \MakeLowercase{\textit{et al.}}: A Sample Article Using IEEEtran.cls for IEEE Journals}

\IEEEpubid{0000--0000/00\$00.00~\copyright~2021 IEEE}

\maketitle

\begin{abstract}
In this work, we investigate the optimal beamformer design for the downlink of Multiple-Input Single-Output (MISO) Non-Orthogonal Multiple Access (NOMA), mainly focusing on a two-user scenario. We derive novel closed-form expressions for the Bit Error Rate (BER) experienced by both users when Quadrature Amplitude Modulation (QAM) is employed. Using these expressions, we formulate a fairness-based optimal beamforming problem aiming to minimize the maximum BER encountered by the users. Due to the complexity of this problem and the time-consuming nature of Constraint Optimization (CO) algorithms for real-time telecommunication systems, we propose a deep learning (DL) approach for its solution. The proposed DL architecture possesses specific input and output characteristics that enable the simultaneous training and use of the system by multiple different antenna schemes. By conducting extensive simulations, we demonstrate that our proposed approach outperforms existing beamforming solutions and achieves BER performance close to that given by CO algorithms while significantly reducing the computational time needed. Finally, we conduct simulations to examine the robustness and efficiency of our system in different test scenarios.

\end{abstract}



\section{INTRODUCTION}

\IEEEPARstart{O}{ne} of the main novelties of fifth-generation (5G) and beyond 5G communications systems is the use of novel Medium Access Schemes (MAC) such as the Non - Orthogonal Multiple Access (NOMA) \cite{dai_non-orthogonal_2015,ding_application_2017}. The key concept introduced by NOMA, and more specifically by power-domain NOMA, is the concurrent access of spectral resources by multiple users and power–domain multiplexing. Power-domain multiplexing is applied by allocating different power levels to different users within the same resource block. In downlink NOMA, the transmitter uses superposition coding after power allocation to combine the signals intended for different users into a single transmission. On the receiver side, each user decodes the combined signal using successive interference cancellation (SIC) and retrieves the transmitted signal designated for them \cite{shin_non-orthogonal_2017}. Following such an approach, we can increase our system's total transmit rate and achieve low latency, massive connectivity, and spectral efficiency.

In the technical literature, various approaches have been proposed to optimize parameters and design in NOMA communication techniques, as highlighted in several studies such as \cite{yang_general_2016,islam_outage_2016,ding_impact_2016}. These studies mainly focus on Single-Input Single-Output (SISO) systems, with \cite{yang_general_2016} concerning NOMA power allocation, \cite{islam_outage_2016} examining outage probability aspects, and \cite{ding_impact_2016} investigating user pairing strategies. Furthermore, subsequent works such as \cite{liu_multiple-antenna-assisted_2018,ali_non-orthogonal_2017,zhou_resource_2018} explore multiple-antenna NOMA scenarios. In \cite{liu_multiple-antenna-assisted_2018}, a comprehensive review of multiple-antenna-aided NOMA structures is provided alongside resource management strategies for Multiple-Input Multiple-Output (MIMO)-NOMA systems. Similarly, \cite{ali_non-orthogonal_2017} explores a multi-user MIMO-NOMA scheme, concentrating on user clustering and power allocation techniques to enhance spectral efficiency. Finally, \cite{zhou_resource_2018} addresses a power minimization problem within a Multiple-Input Single-Output (MISO)-NOMA framework within an energy harvesting context.

However, in their majority, the above works fail to tackle NOMA's most severe limiting factor. That is, it may result in high error rates for the weaker users \cite{ding_unveiling_2020}. Motivated by this, in this work, we focus on optimizing the design of the MISO-NOMA downlink. In more detail, we focus on the case where users are grouped in pairs and seek to solve the optimal beamformer problem in a fashion that maximizes the fairness between them. To this end, we use the maximum Bit Error Rate (BER) metric to express fairness. Besides the fact that this metric introduces fairness in terms of reliability experienced by the users, it also manages to tackle the problem of high error rates discussed earlier. For this reason, the same metric has also been considered in different research works \cite{kara_error_2020,wang_closed-form_2017,garnier_performance_2020,bariah_error_2019,yahya_exact_2021}. Most of these works are dedicated to the application of fairness between two NOMA users, but with system models different to the one that we explore. For example, in \cite{kara_error_2020}, we see a SISO system with the use of Binary Phase Shift Keying (BPSK) modulation that tries to optimize the sole power allocation parameter needed in single antenna NOMA systems, and in \cite{wang_closed-form_2017,garnier_performance_2020} we have similar works that are focused only on Quadrature Phase Shift Keying (QPSK) modulation schemes.

While the maximum BER metric used in this work provides us with a meaningful way to optimize the fairness and reliability of our system, as it will become evident in later sections, it does not allow us to express the optimal beamforming solution in closed form, because it requires solving a non-convex beamforming optimization problem in real-time. To overcome the complications introduced by this fact, in this work, we propose the use of Deep Learning (DL) techniques to learn the association between observed Channel State Information (CSI) and optimal beamforming solution. Our choice is motivated by the fact that DL techniques have shown promising performance results when used to solve optimization problems in communication systems. Thus, they constitute a competitive antagonist to conventional optimization methods. In more detail, in the recent literature, DL has been successfully applied for routing optimization \cite{zhang_deep_2019}, detection and interference management \cite{samuel_deep_2017,yan_signal_2017} and channel estimation \cite{ye_power_2018}. Furthermore, DL has also been used to solve power allocation problems in multi-antenna systems. For example, in \cite{zhao_power_2020,van_chien_power_2020,sanguinetti_deep_2018,wijaya_intercell-interference_2015,sun_learning_2018,gui_deep_2018} DL is used to provide power allocation solutions which maximize the minimum capacity of the users \cite{zhao_power_2020}, maximize the spectral efficiency (SE) \cite{van_chien_power_2020,sanguinetti_deep_2018}, eliminate interference \cite{wijaya_intercell-interference_2015,sun_learning_2018} and optimize the efficiency and reliability of NOMA systems in terms of sum data rate and block error date \cite{gui_deep_2018}.

Motivated by the aforementioned studies and considerations, in this paper, we examine a general MISO-NOMA system and introduce an approach for optimizing the system's beamformers, such as maximizing fairness, measured in terms of BER performance. We base our optimization on the use of DL techniques. The main contributions of this paper are summarized as follows.
\begin{itemize}
    \item We introduce the two-user fairness-based beamformer optimization problem for the downlink of MISO-NOMA. The fairness metric used for this purpose is the maximum BER experienced by the two users. Moreover, we assume that our system respects a total power consumption constraint. For this optimization problem, we show that optimal beamformers can be constructed by restricting ourselves to beamformers exhibiting convenient structures, i.e., beamformers belonging to a specific two-dimensional space, regardless of the number of antennas at the Base Station (BS). As a result, we establish an important dimensionality reduction of the optimal beamforming problem.
    \item Assuming that both users use Quadrature Amplitude Modulation (QAM), we derive new closed-form BER expressions for each user in the MISO-NOMA scenario. To our knowledge, it is the first time closed-form BER expressions for MISO-NOMA appeared in the technical literature. These expressions are used afterward to solve the beamformer design problem. 
    \item We propose a data-driven learning algorithm that is based on the use of DL. We use the channel responses to conduct extensive simulations and provide results and performance analysis of the proposed DL-based method. We demonstrate the efficiency and robustness of the proposed scheme by comparing it with other widely used beamforming techniques. 
\end{itemize}

The rest of the paper is organized as follows. In Sec. II, we present our system model and the formulation of our problem.  Following that, in Sec. III, we offer the BER analysis for our system, and in Sec. IV, we introduce our proposed algorithm. Sec. V presents extensive simulation results, which allow us to evaluate our algorithm's performance and compare it with existing benchmarks. Finally, in Sec. VI, we present the conclusions we derived from our work and the future ideas around it.

\section{SYSTEM MODEL}

\begin{figure*}[t]
\centering
\includegraphics[width=\linewidth]{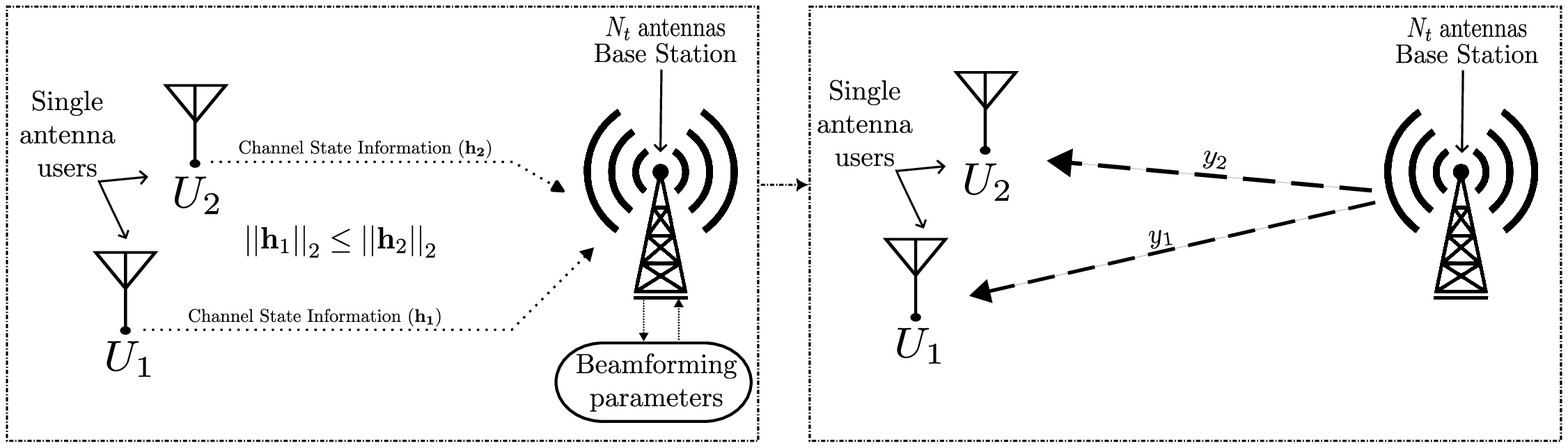}
\caption{System model. The Base Station (BS) is using the perfect Channel State Information (CSI) knowledge in each realization to calculate the optimal beamforming parameters. Afterwards, the downlink transmission begins towards the two NOMA users.}
\label{fig:systemModel_1}
\end{figure*}

We focus on the downlink of the single-cell NOMA-based system shown in Fig. \ref{fig:systemModel_1}, where users are grouped in pairs of two in order to be served concurrently. We note that NOMA system models can generally be implemented with any number of users. However, increasing the number of users will produce significantly more interference, and all the users will suffer from lower Quality of Service (QoS) and error performance \cite{ding_unveiling_2020}. Subsequently, the two-user model is adopted by several recent works \cite{dang_optimal_2023,mouni_short_2023,pendas-recondo_beamforming_2023,mouni_adaptive_2021,xiao_joint_2018,yu_antenna_2018}, as well as, from the 3GPP standards \cite{3gpp.RP.160680}. Our cell's Base Station (BS) is equipped with $N_t$ antennas, and the two users being served concurrently have single-antenna User Equipment (UE). We use the notation $U_n$ for the $n$-th user, $n=1,2$, and we assume that $U_n$ uses QAM, with bits being mapped to QAM symbols using Gray coding. Moreover, we use the notation $M_n$ for the modulation order of the constellation used by $U_n$. Finally, we consider that the BS has perfect knowledge of every channel of the system. Based on this knowledge, the BS will calculate the optimal beamforming parameters and initiate the downlink transmission towards the users. We note that in Table \ref{tab:notations} in Appendix \ref{section:Notations}, we summarize all the significant notations.

\subsection{ OPERATION AT THE TRANSMITTER SIDE (BS)}

Let $s_n$ be the QAM symbol intended for $U_n, n=1,2$ during a given symbol period, and $A_{n,I}, A_{n,Q}$ the in-phase and quadrature components of $s_n$ respectively. It then follows $A_{n,I}$ and $A_{n,Q}$ take values of the form: 
\begin{equation}
A_{n,I} = \pm (2k_n - 1), \;\; \textrm{with}\; \;  k_n = -\frac{\sqrt{M_n}}{2} +1, ...\;, \frac{\sqrt{M_n}}{2},
\end{equation}
and:
\begin{equation}
A_{n,Q} = \pm (2l_n - 1), \; \; \textrm{with} \;\; l_n = -\frac{\sqrt{M_n}}{2} +1, ...\;, \frac{\sqrt{M_n}}{2},
\end{equation}
and the transmitted QAM symbol is expressed as:
\begin{equation}
s_{n} = \frac{A_{n,I}}{\sqrt{E_n}} + j \frac{A_{n,Q}}{\sqrt{E_n}},
\end{equation}
where $E_n = \frac{2}{3}(M_n - 1) ,n=1,2$ are normalization constants used such as to ensure that the average power of the constellation used by the two users is equal to one \cite{proakis_digital_2008}.

In order to properly exploit the multiple antennas at the BS, the BS applies beamforming on the signal of the $n$-th user, $n=1,2,$ before the transmission. As a result, the signal which is transmitted is expressed as:
\begin{equation}
\mathbf{x} = \mathbf{w}_1s_1 + \mathbf{w}_2s_2 
\label{eq:x}
\end{equation}
where  $\mathbf{w}_n \in \mathbb{C}^{N_t} $ the complex beamforming vector for user $U_n$, with $\left\|\mathbf{w}_n \right\|_2=1, n=1,2$. Given the above signal model at the transmitter side, we describe the operation at the receiver side in what follows.

\subsection{OPERATION AT THE RECEIVER SIDE (USERS)}

Given that both users employ single-antenna terminals, the signal at $U_n$ is expressed as:
\begin{equation} 
y_n = \mathbf{h}_n^H\mathbf{x} + \nu_n.
\label{eq:y_n}
\end{equation}
where $\nu_n \sim \mathcal{CN} \left(0, N_0  \right)$ is the additive white Gaussian noise at the $n$-th user, with $N_0$ being the noise variance. Moreover, vector $\mathbf{h}_n \in \mathbb{C}^{N_t}$ is the channel vector formed between the BS and $U_n$. We note that the BS has perfect CSI knowledge. We model the channel vector using a combination of path loss and independent identically distributed (i.i.d.) Rayleigh fading channel. As a result, $\mathbf{h}_n$ is expressed as:
\begin{equation}
\mathbf{h}_n = \sqrt{\beta_n}\mathbf{\tilde{h}}_n
\label{kanaliRayleigh}
\end{equation}
where $\mathbf{\tilde{h}}_n \sim \mathcal{CN}(\mathbf{0}, \mathbf{I})$ represents the small-scale fading and $\beta_n$ represents the large-scale fading that can be formulated as

\begin{equation}
\beta_n = 10^{\frac{\textrm{PL}_n}{10}}
\end{equation}
with $\textrm{PL}_n$ the path loss expressed in dB as in \cite{ngo_cell-free_2017}.

As far as message decoding is concerned, we assume that $\left\| \mathbf{h}_1\right\|_2 \leq \left\| \mathbf{h}_2 \right\|_2$ and that, as in \cite{8411153,choi_generalized_2017} the user with the weakest channel (i.e., $U_1$) directly decodes the signal intended for it. In more detail, the decoding process at the two users is summarized as follows.

\subsubsection{Decoding at $U_1$}
Combining \eqref{eq:x} and \eqref{eq:y_n} the signal received by $U_1$ can be written as:
\begin{equation}
y_1 = \mathbf{h}_1^H\mathbf{w}_1s_1 + \mathbf{h}_1^H\mathbf{w}_2s_2 +\nu_1.
\end{equation}
As a result, assuming perfect knowledge of the channel responses and the beamformers applied by the BS, $U_1$ starts by applying phase correction, such as to correct the rotation caused by the channel to its constellation. This phase correction involves multiplying $y_1$ with a complex exponential of phase selected to be opposite to the phase of the complex quantity $\mathbf{h}_1^H\mathbf{w}_1$. In more detail, introducing terms
\begin{equation}
\theta_{1,j} = 
\angle \mathbf{h}_1^H\mathbf{w}_j, \;\; j=1,2,
\end{equation}
phase correction at $U_1$ involves multiplying $y_1$ with $e^{-j \theta_{1,1}}$. The resulting signal is then expressed as:
\begin{equation}
\label{eq:tilde_y_1}
\tilde{y}_1 = |\mathbf{h}_1^H\mathbf{w}_1|s_1 + |\mathbf{h}_1^H\mathbf{w}_2|
e^{j\left(\theta_{1,2} - \theta_{1,1} \right)}s_2 + \tilde{\nu}_1
\end{equation}
where $\tilde{\nu}_1 = \nu_1e^{-j\theta_{1,1}} $.
Since no interference cancellation is applied at $U_1$, the detection of $s_1$ at $U_1$ is based on $\tilde{y}_1$. Moreover, for the detection process, we consider the use of a maximum likelihood (ML) detector and assume perfect knowledge of the channel coefficient $\hat{\mathbf{h}}_1 = \mathbf{h}_1$. The detection decision is then expressed as:
\begin{equation}
\hat{s}_1 = \arg\min\limits_{s_1}\bigg|\tilde{y}_1 - |\mathbf{h}_1^H\mathbf{w}_1|s_1\bigg|^2.
\label{eq:hat_s_1FIRST}
\end{equation}

\subsubsection{Decoding at $U_2$}
Using again \eqref{eq:x} and \eqref{eq:y_n}, the received signal at $U_2$ can be written as 
\begin{equation}
y_2 = \mathbf{h}_2^H\mathbf{w}_1s_1 + \mathbf{h}_2^H\mathbf{w}_2s_2 +\nu_2.
\end{equation}
Using $y_2$, $U_2$ first applies SIC to decode and remove the signal of $U_1$ from the received one.  Following that, it decodes the signal that was intended for it. This process is described in more detail in what follows.

\paragraph{SIC at $U_2$}
In order to apply SIC, $U_2$ first corrects any phase rotation caused by the channel and the beamforming process to the signal of $U_1$. To describe this rotation, as well as the SIC process, let us introduce the variables:
\begin{equation}
\theta_{2,j} =\angle \mathbf{h}_2^H\mathbf{w}_j, \;\; j=1,2.
\end{equation}
We can then express the received signal at $U_2$ as:
\begin{equation}
\label{eq:tilde_y_2_BER}
\tilde{y}_2 = |\mathbf{h}_2^H\mathbf{w}_1|e^{j(\theta_{2,1} - \theta_{2,2})}s_1 + |\mathbf{h}_2^H\mathbf{w}_2|s_2 + \tilde{\nu}_2,
\end{equation}
with $\tilde{\nu}_2 = \nu_2e^{-j\theta_{2,2}} $. Using $\tilde{y}_2$, the detection of $U_1$'s signal at $U_2$ can be mathematically described through the operation:
\begin{equation}
    s'_1 = \arg\min\limits_{s_1}\bigg|\tilde{y}_2 - |\mathbf{h}_2^H\mathbf{w}_1|e^{j(\theta_{2,1} - \theta_{2,2})}s_1 \bigg|^2.
\label{eq:hat_s_2SICFIRST}
\end{equation}
By using $s'_1$, $U_2$ performs SIC such as to detect the signal intended for it. This is done as follows.

\paragraph{Detecting $s_2$ at $U_2$}
After using $s'_1$ such as to perform SIC, the detection of signal $s_2$ at $U_2$ is mathematically described by the following operation:
\begin{equation}
\hat{s}_2 = \arg\min\limits_{s_2}\bigg|\tilde{y}_2 - |\mathbf{h}_2^H\mathbf{w}_1|e^{j(\theta_{2,1} - \theta_{2,2})}s'_1 - |\mathbf{h}_2^H\mathbf{w}_2|s_2\bigg|^2.
\label{eq:hat_s_2FIRST}
\end{equation}
From \eqref{eq:hat_s_2FIRST}, we can directly see that in case that $s_1' \neq s_1$ the impact of errors during the detection of $s_1$ can impact the detection of $s_2$. This impact will be further quantified in the next sections of the paper, where a novel BER expression for the BER of user $U_2$ is derived. Having defined our system operation, we now introduce the performance optimization problem of interest for our system design.

\subsection{THE CONSIDERED OPTIMIZATION PROBLEM}
Working on the system model introduced above, in this paper, we seek to optimize the downlink beamformer design at the BS such as to maximize the fairness of our system, subject to a total power consumption constraint. To this end, we set the minimization of the maximum BER of the two users as the objective of our system design. In more detail, the system design that we consider is based on solving the following optimization problem:
\begin{equation}
\begin{split}
{}&{}
\textrm{minimize}_{\mathbf{w}_1,\mathbf{w}_2}\{\max\{P_{e}^{(1)} ,P_{e}^{(2)} \}\} \\&
\textrm{subject to:}\;\;
\left\| \mathbf{w}_1\right\|^2 +
\left\| \mathbf{w}_2\right\|^2 \leq P_{max},
\end{split}
\label{argminmax}
\end{equation}
where $P_{e}^{(n)},n=1,2,$ is the BER experienced by $U_n$ and $P_{max}$ indicates the total power consumption. The reason for choosing a BER-focused performance metric is that such a metric allows the capture of one of the most critical physical layer (PHY) characteristics of a communication system, which is the modulation scheme. As a result, using such a performance metric allows optimizing practical communications systems and the implementation of already existing standardized PHY solutions. This is not necessarily the case when rate-based metrics, such as the sum rate, are used, which cannot capture fundamental network characteristics (such as the modulation scheme) and the impact of imperfect SIC when practical modulation schemes like QAM are employed at the PHY. BER, on the other hand, quantifies the likelihood of individual bit errors in data transmission. Therefore, the BER allows for a more precise assessment of the system performance, which considers not only whether the signal is lost, like in the case that outage probability is used as a metric, but also the severity of data corruption.

In the following sections, we discuss the process of solving problem \eqref{argminmax}. This process also includes the derivation of novel closed-form expressions for the BER of the two users.

\section{BER ANALYSIS OF OUR SYSTEM}
In this section, we take the first steps to solve problem \eqref{argminmax}. These include
\begin{inparaenum}
\item the derivation of a new result concerning the structure of the optimal beamformer for the considered problem. This result allows us to greatly simplify the problem of optimal beamforming as well as to limit the search space for the optimal beamformers $\mathbf{w}_1$ and $\mathbf{w}_2$ to a two-dimensional space, regardless of the number of transmit antennas at the BS.
\item the derivation of novel exact expressions for the BER experienced by the two users. The main novelty of these expressions is that they account for the fact that, as a result of the beamforming process, the signals of the two users arrive at the receiver having experienced different phase rotations. As a result of the different phase rotations, the standard BER expressions of SISO NOMA systems are not applicable. Finally, we note that to the best of our knowledge, our closed-form BER expressions are the first to appear in the literature and consider the effects of MISO channels in NOMA transmission.
\end{inparaenum}
In what follows, we further elaborate on the two aforementioned items, focusing initially on the topic of determining the structure of the optimal beamformer.

\subsection{DETERMINING A SIMPLE STRUCTURE FOR THE OPTIMAL BEAMFORMER}

\label{sec:ber_analysis}
In order to derive a simple structure for the optimal beamformer, we use the approach introduced in \cite{ropokis_multi-relay_2019} and \cite{demir_optimum_2018} and express beamformers $\mathbf{w}_1$ and $\mathbf{w}_2$ with the aid of an orthonormal basis $\{ \mathbf{u}_1,...,\mathbf{u}_N  \}$ for $\mathcal{C}^{N_t \times 1}$, where we have selected vectors $\mathbf{u_1}$ and $\mathbf{u_2}$ as follows:
\begin{equation}
\mathbf{u_1} = \frac{\mathbf{h_1}}{\lVert \mathbf{h_1} \rVert}, \;\;\;\;\; \mathbf{u_2} = \frac{(\mathbf{I} - \mathbf{u_1u_1^H})\mathbf{h_2}}{\lVert  (\mathbf{I} - \mathbf{u_1u_1^H})\mathbf{h_2}   \rVert}.
\label{orth_basis}
\end{equation}
Beamformers $\mathbf{w}_n, n=1,2,$ can then be written in the form
\begin{equation}
\mathbf{w}_1 = \sum_{i = 1}^{N}\rho_ie^{j\theta_i}\mathbf{u}_i, \;\;
\textrm{and} \;\;
\mathbf{w}_2 = \sum_{i = 1}^{N}\delta_ie^{j\phi_i}\mathbf{u}_i
\label{eq:w_1_w_2}.
\end{equation}
where $\rho_i$ and $\delta_i \in \mathbb{R}^{+}, i=1, \ldots, N$, and $\theta_i, \phi_i \in [ 0 , 2\pi), i=1,\ldots, N$. Using this representation for the beamformers, the following two theorems help us reformulate the decision variables $\tilde{y}_1$ in \eqref{eq:tilde_y_1} and $\tilde{y}_2$ in \eqref{eq:tilde_y_2_BER}
which are used for detection purposes at $U_1$ and $U_2$ respectively.
\begin{theorem}
\label{theor:1}
The decision variable $\tilde{y}_1$ in \eqref{eq:tilde_y_1} can be equivalently written as:
\begin{equation}\label{eq:tilde_y_1_BEAM}
\tilde{y}_1 = \rho_1\lVert \mathbf{h}_1 \rVert s_1 +\delta_1e^{j( \tau_1)}\lVert \mathbf{h}_1 \rVert s_2 + \tilde{\nu}_1
\end{equation}
where $ \tau_1 = \phi_1 - \theta_1$.

\end{theorem}

\begin{IEEEproof}
Using the aforementioned orthonormal basis, we obtain: 
\begin{equation}
\label{eq:inner_products}
\mathbf{h}_1^H\mathbf{u}_1 =  \lVert \mathbf{h_1} \rVert, \;\;
\textrm{and}\;\;
\mathbf{h}_1^H\mathbf{u}_m = \lVert \mathbf{h}_1 \rVert\mathbf{u}^H_1\mathbf{u}_m,
\end{equation}
for $m=2,\ldots, N$.
Hence, by substituting \eqref{eq:w_1_w_2} in \eqref{eq:tilde_y_1} and exploiting \eqref{eq:inner_products} as well as the fact that $\mathbf{u}_1^{H} \mathbf{u}_m =0$, for $m\neq 1$ the form described by the theorem for $\tilde{y}_1$ is obtained.
\end{IEEEproof}

While Theorem \ref{theor:1} gives a convenient expression for $\tilde{y}_1$ the following one provides a convenient expression for $\tilde{y}_2$.
\begin{theorem}
\label{theor:2}
The decision variable $\tilde{y}_2$ can be equivalently written as:
\begin{equation}
\begin{split}
{}&{}
\tilde{y}_2 =\big(\rho_1\big | \mathbf{h}_2^H\mathbf{u}_1\big | +\rho_2\big |\mathbf{h}_2^H\mathbf{u}_2 \big |\big)s_1\\&
\;\;\;\;+ \big|\delta_1e^{j\phi_1}\mathbf{h}_2^H\mathbf{u}_1 + \delta_2e^{j\phi_2}\mathbf{h}_2^H\mathbf{u}_2\big|s_2 + \tilde{\nu}_2.
\label{eq:eq:tilde_y_2_BEAM}
\end{split}
\end{equation}
\end{theorem}
\begin{IEEEproof}
We start by noticing that due to the construction of the orthonormal basis $\left\{ \mathbf{u}_1, \ldots, \mathbf{u}_n \right\}$, it is easy to prove that $h_2^H\mathbf{u}_m = 0$ for $m = 3,...,N$. Using this fact along with \eqref{eq:w_1_w_2}, the signal which is going to be used at $U_2$ in order to extract the signal intended for $U_1$ is written as:
\begin{equation}
\begin{split}
{}&{}
\tilde{y}_2 =\big(\rho_1e^{j\theta_1}\mathbf{h}_2^H\mathbf{u}_1 +\rho_2e^{j\theta_2}\mathbf{h}_2^H\mathbf{u}_2 \big)s_1 \cdot e^{-j\hat{\phi}} \\&
\;\;\;\;+ \big|\delta_1e^{j\phi_1}\mathbf{h}_2^H\mathbf{u}_1 + \delta_2e^{j\phi_2}\mathbf{h}_2^H\mathbf{u}_2\big|s_2 + \tilde{\nu}_2,
\end{split}
\end{equation}
where $\hat{\phi} = \angle\big(\delta_1e^{j\phi_1}\mathbf{h}_2^H\mathbf{u}_1 + \delta_2e^{j\phi_2}\mathbf{h}_2^H\mathbf{u}_2\big)$.
As a result, in order to maximize the probabilities of correct SIC, the two components of the beamformer concerning symbol $s_1$ should be aligned such as to ensure that:
\begin{equation}\label{klasma_1}
\frac{e^{j\theta_1}\mathbf{h}_2^H\mathbf{u}_1}{|\mathbf{h}_2^H\mathbf{u}_1|} = \frac{e^{j\theta_2}\mathbf{h}_2^H\mathbf{u}_2}{|\mathbf{h}_2^H\mathbf{u}_2|}= e^{j\hat{\phi}}.
\end{equation}
The result of \eqref{eq:eq:tilde_y_2_BEAM} is therefore obtained.
\end{IEEEproof}

Combining the results of Theorems \ref{theor:1} and \ref{theor:2} the following Lemma is derived.
\begin{lemma}
\label{lemma:1}
An optimal beamformer pair $\mathbf{w}_1^{\star}, \mathbf{w}_2^{\star}$ for problem \eqref{argminmax} can be found where $\mathbf{w}_1^{\star}$ is written as:
\begin{equation}
\mathbf{w}_1^{\star} =
\rho_1^{\star}  e^{j \theta_1^{\star}} \mathbf{u}_1 + 
\rho_2^{\star} e^{j \theta_2^{\star}} \mathbf{u}_2,
\end{equation}
and $\mathbf{w}_2^{\star}$ as:
\begin{equation}
\mathbf{w}_2^{\star} =
\delta_1^{\star}  e^{j \phi_1^{\star}} \mathbf{u}_1 + 
\delta_2^{\star} e^{j \phi_2^{\star}} \mathbf{u}_2.
\end{equation}
\end{lemma}
\begin{IEEEproof}

Starting from the results of Theorems \ref{theor:1} and \ref{theor:2}, we directly  see that  decision variables $\tilde{y}_1$ and $\tilde{y}_2$ only depend on $\delta_i, \rho_i, \phi_i, \theta_i, i=1,2$. As a result, only these quantities influence the BER of the two users while the values of  $\delta_i, \rho_i, \phi_i, \theta_i, i = 3, \ldots, N$  do not appear in the cost function of the problem of minimizing the maximum BER ($\Psi$). Moreover, by rewriting the constraint $\left\| \mathbf{w}_1 \right\|^2 + \left| \mathbf{w}_2\right|^2 \leq P_{max}$ as $\sum_{i=1^N} \rho_i^2 + \delta_i^2 \leq P_{max}$ we directly see that non-zero values for parameters $\delta_i$ and $\rho_i$ for $i= 3, \ldots, N$ only restrict the feasible values for $\rho_i, \delta_i, i=1,2$. Hence, by setting $\rho_i=0$ and $\delta_i=0$ for $i\geq3$, we maximize the search space for variables $\rho_i, \delta_i, i=1,2$. The result of the Lemma then follows.
 \end{IEEEproof}

The result of Lemma \ref{lemma:1} clearly ensures the existence of a simple beamformer structure for the optimal beamforming solution. However, to proceed further and solve \eqref{argminmax}, the derivation of a convenient closed-form expression for its cost function is required. Thus, we derive novel expressions for the BER experienced by $U_1$ and $U_2$.

\subsection{BER PERFORMANCE AT $U_1$}
Using the result of Theorem \ref{theor:1} and \eqref{eq:tilde_y_1_BEAM}, the ML decoder in \eqref{eq:hat_s_1FIRST} is written as:
\begin{equation}
\hat{s}_1 = \arg\min\limits_{s_1}\bigg|\tilde{y}_1 - \rho_1\lVert \mathbf{h}_1 \rVert s_1\bigg|^2.
\label{eq:hat_s_1}
\end{equation}
In what follows, using \eqref{eq:hat_s_1}, we derive a closed-form expression for the BER of $U_1$. Let $P_{b,i}^{(1)}$ be the probability of an error on the $i-th$ bit of a symbol of $U_1$. As discussed in \cite{assaf_exact_2020}, one can then express the average conditional BER for $U_1$ as:
\begin{equation}
P_{e}^{(1)} = 
\frac{2}{\log_2\sqrt{M_1}}
\sum_{i=1}^{\log_2\sqrt{M_1}} P_{b,i}^{(1)}.
\end{equation}
and the BER expression can be derived by finding the probability of error for each one of the transmit bits which constitute a QAM symbol, separately. Let us therefore focus on the $i$-th bit of a symbol, $i \in \left\{1,2, \ldots, \log_2 M_1 \right\}$. In the following Theorem, we present our novel result, which provides us with a closed form for $P_{b,i}^{(1)}$.
\begin{theorem}
Assuming Gray coding, the BER for the $i$-th bit of a QAM symbol is given as:
\begin{equation}
\begin{split}
{}&{}
P_{b,i}^{(1)} =\frac{1}{M_2\sqrt{M_1}} \times 
\sum_{k=0}^{L_{1,1}}\sum_{l=0}^{\Lambda_2}\sum_{m=0}^{\Lambda_2}\Bigg[D_1(i,k) \cdot \\&
\;\;\;\;\;\;\;\;\;\;\;Q\bigg(g_1 \bigg(2k+1,2l-\sqrt{M_2}+1,2m-\sqrt{M_2}+1\bigg)\bigg) \Bigg],\\&
\end{split}
\label{eq:P_bi_1_expr_1}
\end{equation}
where
\begin{equation}
\begin{split}
{}&{}
D_1(i,k) = (-1)^{\lambda_{i,k,1}} \bigg( 2^{i-1} -   \bigg\lfloor \frac{k2^{i-1}}{\sqrt{M_1}} + \frac{1}{2} \bigg\rfloor \bigg), \\& \;\;\;\;\;\;\;\;\;\;\;
L_{1,1} = (1-2^{-i})\sqrt{M_1}-1, \\& \;\;\;\;\;\;\;\;\;\;\;
\lambda_{i,k,1} =\bigg \lfloor \frac{k2^{i-1}}{\sqrt{M_1}} \bigg \rfloor, \\& \;\;\;\;\;\;\;\;\;\;\; \Lambda_n = \sqrt{M_n} - 1
\end{split}
\label{eq:P_bi_1_expr_2}
\end{equation}
and
\begin{equation}
\begin{split}
g_1 \left( a, b, c \right) = 
\frac{\frac{a \cdot\rho_1\lVert \mathbf{h}_1 \rVert}{\sqrt{E_1}} +\frac{\delta_1\lVert \mathbf{h}_1 \rVert}{\sqrt{E_2}}\bigg(b \cdot \cos\tau_1  + c \cdot \sin\tau_1 \bigg)}{\sqrt{N_0/2}},\\&
\end{split}
\label{eq:P_bi_1_expr_3}
\end{equation}
with $Q(\cdot)$ being the $Q$-function.
\end{theorem}
\begin{IEEEproof}
We notice that due to Gray coding, demodulation of the first $\log_2\sqrt{M_1}$ bits of $U_1$ can be done using only the real part of $\tilde{y}_1$ and the demodulation of the last $\log_2\sqrt{M_1}$ bits can be done using the imaginary part of $\tilde{y}_1$. As a result, the interference experienced during demodulation of any of the bits of $U_1$ carried by the real part of $\tilde{y}_1$ is equal to:
\begin{equation}
I_{real,1} =
\Re \left\{
\delta_1e^{j( \tau_1)}\lVert \mathbf{h}_1 \rVert s_2 + \tilde{\nu}_1 \right\}
\label{eq:realInter1}
\end{equation}
and the interference experienced during demodulation of any of the bits of $U_1$ carried by the imaginary part of $\tilde{y}_1$ is equal to:
\begin{equation}
I_{imag,1} =
\Im \left\{
\delta_1e^{j( \tau_1)}\lVert \mathbf{h}_1 \rVert s_2 + \tilde{\nu}_1 \right\}.
\label{eq:imagInter1}
\end{equation}
Unlike \cite{assaf_exact_2020}, the terms in \eqref{eq:realInter1} and \eqref{eq:imagInter1} include interference both from the real and the imaginary parts of $s_2$. Thus, extending the BER expression from SISO to MISO systems is not a straightforward procedure. In our analysis, these interference terms and their effect are captured by the triple sum in \eqref{eq:P_bi_1_expr_1}. Moreover, terms $D_1, L_{1,1}$ and $\Lambda_n$ provide a compact way to shape the expression $P_{b,i}^{(1)}$, as they indicate the number of $Q$-functions that are needed for each QAM modulation pair ($M_1$, $M_2$). Further details on the derivation can be found in Appendix \ref{section:Appendix}, where this process is further explained for the example of 4-QAM constellation use by both users.
\end{IEEEproof}

Having presented the BER expression for $U_1$, we now discuss the BER expression for $U_2$.

\subsection{BER PERFORMANCE AT $U_2$}

As mentioned before, with the use of the detector in $U_2$ we want to decode and remove $U_1$'s symbol from the received signal and then decode the symbol intended for $U_2$. Consequently, with the help of \eqref{eq:hat_s_2FIRST} and \eqref{eq:eq:tilde_y_2_BEAM} the detector for $U_2$'s signal will be

\begin{equation}
\begin{split}
{}&{}
\hat{s}_2 = \arg\min\limits_{s_2}\bigg|\tilde{y}_2 - \big(\rho_1\big | \mathbf{h}_2^H\mathbf{u}_1\big | +\rho_2\big |\mathbf{h}_2^H\mathbf{u}_2 \big |\big)s'_1 \\& \;\;\;\;\;\;\;\;\;\;\;\;\;\;\;\;\;\;\;\;-\big|\delta_1e^{j\phi_1}\mathbf{h}_2^H\mathbf{u}_1 + \delta_2e^{j\phi_2}\mathbf{h}_2^H\mathbf{u}_2\big|s_2\bigg|^2,
\end{split}
\label{eq:hat_s_2}
\end{equation}
where $s'_1$ is the estimated $U_1$'s signal at the second user and can be presented as
\begin{equation}
    s'_1 = \arg\min\limits_{s_1}\bigg|\tilde{y}_2 -\big(\rho_1\big | \mathbf{h}_2^H\mathbf{u}_1\big | +\rho_2\big |\mathbf{h}_2^H\mathbf{u}_2 \big |\big)s_1 \bigg|^2.
\label{eq:hat_s_2SIC}
\end{equation}

Based on the nature of QAM constellations and the use of Gray coding of the symbols at each user, a pattern emerges, similar to the one that appears for $U_1$. As a result, the BER can be derived by separating the bits of $U_2$ into two equal groups, each consisting of $\log_2\sqrt{M_2}$ bits, and the two groups have the same BER. Consequently, the average conditional BER expression for $U_2$ will be:
\begin{equation}
P_{e}^{(2)} = \frac{2}{\log_2\sqrt{M_2}}\sum_{i=1}^{\log_2\sqrt{M_2}}{P_{b,i}^{(2)}}.
\end{equation}
Again, the derivation of this BER expression can be done by finding the probability of error for each one of the transmitted bits which constitute a QAM symbol, separately. By focusing on the $i$-th bit of a symbol, where $i \in \left\{1,2, \ldots, \log_2 M_2 \right\}$, the expression for $P_{b,i}^{(1)}$ is given from the following theorem.

\begin{theorem}
Assuming Gray coding, the BER for the $i$-th bit of a QAM symbol is given as:
\begin{equation}
\begin{split}
{}&{}
P_{b,i}^{(2)} =\frac{1}{\sqrt{M_1 \cdot M_2}} \times \\&
\;\;\;\;\;\;\;\;\;
\bigg(\sum_{k=0}^{L_{1,2}}\sum_{l=0}^{2\Lambda_1}SD_2(i,k)D_3(i,l)Q\bigg(g_{2,i}^+\bigg(l,2k+1\bigg)\bigg)\\&
\;\;\;\;\;
-  \sum_{k=0}^{L_{1,2}}\sum_{l=1}^{2\Lambda_1}SD_2(i,k)D_3(i,l)Q\bigg(g_{2,i}^-\bigg(l,2k+1\bigg) \bigg)\bigg)\\&
\textrm{where}\;
S = (-1)^{ \big \lfloor \frac{l2^{(\log_2\sqrt{M_2})+i-1}}{\sqrt{M_2}} \big \rfloor + \lambda_{i,k,2} }, \\&
\;\;\;\;\;\;\;\;\;\;
 \Lambda_n = \sqrt{M_n} - 1, \\&
 \;\;\;\;\;\;\;\;\;\; D_2(i,k) = 2^{i-1} - \bigg \lfloor \frac{k2^{i-1}}{\sqrt{M_2}} + \frac{1}{2} \bigg \rfloor, \\&
\;\;\;\;\;\;\;\;\;\;
D_3(i,l) = 2^{\log_2\sqrt{M_1}} - \bigg \lfloor \frac{l}{ 2^{1-(i-1) \log_2(\sqrt{M_1}-1)}} + \frac{1}{2} \bigg \rfloor,\\& \;\;\;\;\;\;\;\;\;\;
L_{1,2} = (1-2^{-i})\sqrt{M_2}-1, \;\;
\lambda_{i,k,2} =\bigg \lfloor \frac{k2^{i-1}}{\sqrt{M_2}} \bigg \rfloor,\\&
\textrm{and}\\&
g_{2}^\pm \left( a, b \right) = 
\frac{a \frac{\big(\rho_1\big | \mathbf{h}_2^H\mathbf{u}_1\big | +\rho_2\big |\mathbf{h}_2^H\mathbf{u}_2 \big |\big)}{\sqrt{E_1}} \pm b \frac{\big|\delta_1e^{j\phi_1}\mathbf{h}_2^H\mathbf{u}_1 + \delta_2e^{j\phi_2}\mathbf{h}_2^H\mathbf{u}_2\big|}{\sqrt{E_2}}}
{\sqrt{N_0/2}}
\end{split}
\label{eq:P_bi_2_exprFINAL}
\end{equation}
with $Q(\cdot)$ being the $Q$-function.
\end{theorem}

\begin{figure}[h!]
\includegraphics[width=\linewidth]{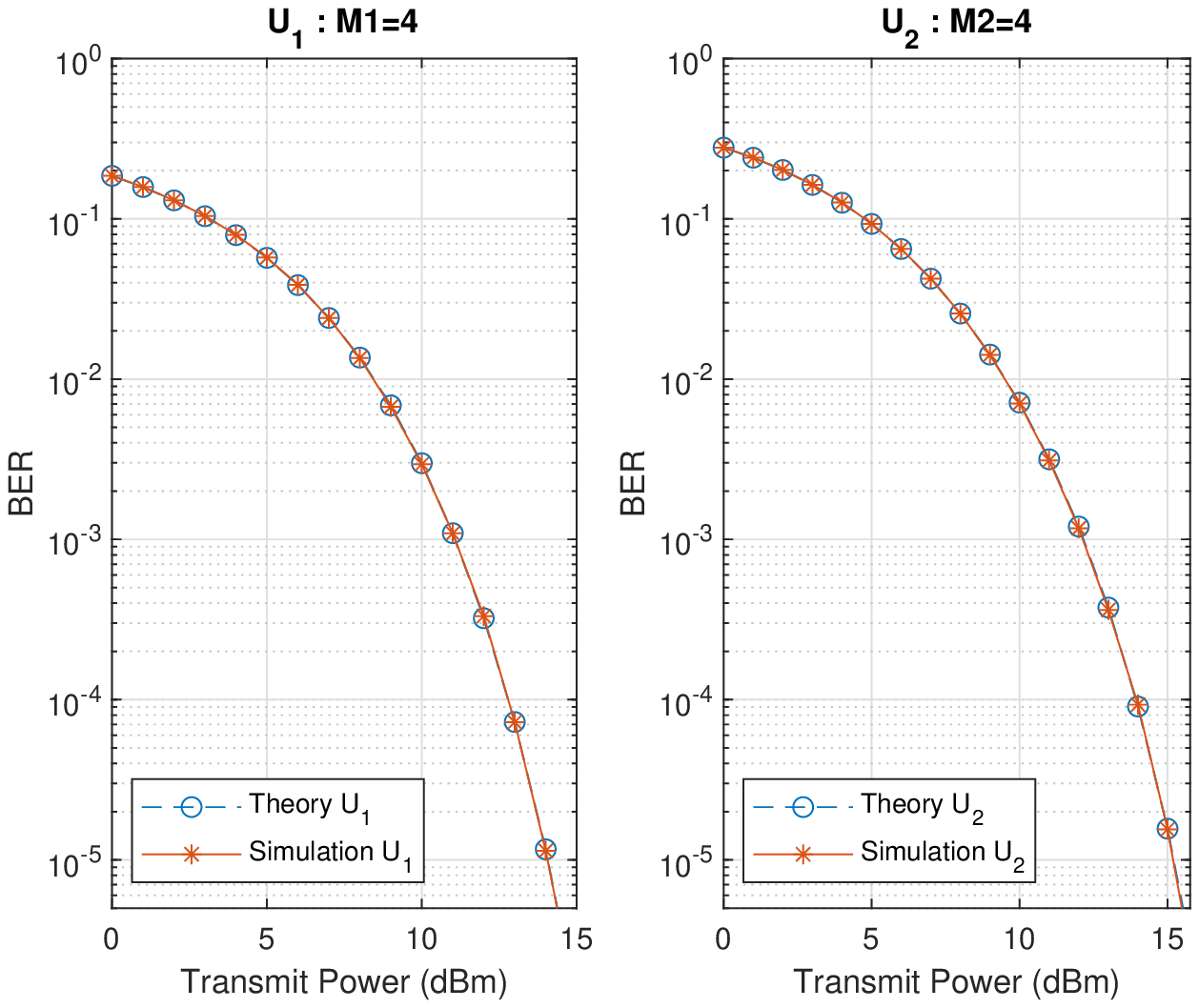}
\caption{Average BER using $M_1 = M_2$ for $U_1$ and $U_2$.}
\label{fig:avBerappend}
\end{figure}

\begin{IEEEproof}
As mentioned before, the probability of error for $U_2$ depends on $s'_1$ which is the estimation of $U_1$'s symbol during the SIC. Due to the modulation's nature, the error probability of each bit of $U_2$'s symbol is connected only with specific bits of the first user's estimated symbol. The bits of $U_1$'s symbol can be divided into two equal groups based on the nature of interference that they cause on the received signal. The first group can be the set of symbols that introduce interference at the real part of the received signal, and the second group can contain the bits that introduce interference at the imaginary part of the signal. Each group of $U_1$'s bits can affect $\log_2 \sqrt{M_1}$ bits of the second user's symbol. Thus, the interference that is experienced during the demodulation of the $s_2$ symbol can be expressed as
\begin{equation}
I_{real,2} =
\Re \left\{
\big(\rho_1\big | \mathbf{h}_2^H\mathbf{u}_1\big | +\rho_2\big |\mathbf{h}_2^H\mathbf{u}_2 \big |\big)s'_1 + \tilde{\nu}_2 \right\},
\end{equation}
for the bits of $s_2$ that are carried by the real part of $\tilde{y}_2$, and as 
\begin{equation}
I_{imag,2} =
\Im \left\{
\big(\rho_1\big | \mathbf{h}_2^H\mathbf{u}_1\big | +\rho_2\big |\mathbf{h}_2^H\mathbf{u}_2 \big |\big)s'_1 + \tilde{\nu}_2 \right\}
\end{equation}
for the bits of $s_2$ that are carried by the imaginary part of $\tilde{y}_2$. Expressions $S, D_2,D_3,L_{1,2}$ and $\Lambda_n$ define the number of $Q$-functions that will appear in each part of $P_{b,i}^{(2)}$ expression.
\end{IEEEproof}

In Appendix \ref{section:Appendix}, we present the case of $M_1 = M_2 = 4$ for QAM constellations in detail. Additionally, in Fig. \ref{fig:avBerappend}, we can observe the analytical and the simulated BER of both users for the aforementioned case.

\subsection{BER OPTIMIZATION PROBLEM}

Based on the results of the previous analysis, the search for beamforming vectors that provide the lowest possible $\Psi$ between the two users (\ref{argminmax}), comes down to finding the optimal $\rho_i,\delta_i, \tau_i, \phi_i$ parameters and can be formulated as

\begin{equation} \label{newargminmax}
\begin{split}
{}&{}
\textrm{minimize}_{\rho_1,\rho_2,\delta_1,\delta_2, \tau_1,\phi_1,\phi_2   }\{\max\{P_{e}^{(1)} ,P_{e}^{(2)} \}\} \\&
\textrm{subject to:} \;\; \eqref{constr1},\;\eqref{constr2},\;\eqref{constr3}.
\end{split}
\end{equation} 
We note that we use only three phase parameters for our optimization ($\tau_1,\phi_1,\phi_2$) due to the phase shifting that occurs in \eqref{eq:tilde_y_1_BEAM} and in \eqref{eq:eq:tilde_y_2_BEAM} that eliminates the extra phase parameters.

To ensure correct transmission and prevent any overlap among users' symbols, it is necessary for the beamformer parameters to adhere to the following constraints. As we can observe in \cite{lee_average_2019}, the following rate has to be satisfied
\begin{equation}
\bigg|\frac{ \rho_1}{\delta_1(\cos(\tau_1) - \sin(\tau_1)) \cdot \Lambda_2 } \bigg| > Y_1,
\label{constr1}
\end{equation}
where $Y_1 = \frac{M_1 -1}{M_2 - 1} \cdot (\sqrt{M_2}-1)^2 $. \\

Likewise, by following the analysis of the previous expressions, we reach the conclusion that in order for the SIC to be reliable, the following expression also has to be valid 
\begin{equation}
\frac{\rho_1\big | \mathbf{h}_2^H\mathbf{u}_1\big | +\rho_2\big |\mathbf{h}_2^H\mathbf{u}_2 \big |}{\big|\delta_1e^{j\phi_1}\mathbf{h}_2^H\mathbf{u}_1 + \delta_2e^{j\phi_2}\mathbf{h}_2^H\mathbf{u}_2\big|}  > Y_1.
\label{constr2}
\end{equation}
In addition to these restrictions, we have to consider that there are limitations on the maximum transmit power of our system. By normalizing the transmit power of the constraints in  (\ref{argminmax}) to unity, we derive the following expression
\begin{equation}
\rho_1^2 + \rho_2^2 + \delta_1^2 + \delta_2^2 \leq 1.
\label{constr3}
\end{equation}

It has to be noted that the objective function of our problem \eqref{newargminmax} is to find the maximum between two expressions that contain multiple $Q$-function terms. It is known that the $Q$-function is neither convex nor concave. Moreover, expressions $P_{b,i}^{(1)}$ in \eqref{eq:P_bi_1_expr_1} and $P_{b,i}^{(2)}$ in \eqref{eq:P_bi_2_exprFINAL} involve the sum and the difference of sums of non-convex functions respectively. As a result, there is no guarantee that problem \eqref{newargminmax} is convex, and generally, non-convexity of the problem should be assumed. Thus, the problem cannot be solved with conventional convex optimization techniques like the ones found in \cite{boyd_convex_2015}. Additionally, given the form of $P_{e}^{(1)}$ and $P_e^{(2)}$, which is complicated and (especially for the case of high-order QAM) involves the summation of a large number of terms the calculation of the cost function of \eqref{newargminmax} can be expensive in practice and the use of standard approaches may not be possible.

Therefore, in an effort to create a fast and robust system that can predict the optimal beamformer that achieves fairness between the users in terms of conditional BER, we apply Deep Learning techniques. Specifically, we use Neural Networks to learn the non-linear correlation that exists between the CSI of the two users and the characteristics of their beamformers.

\section{PROPOSED LEARNING ALGORITHM}

In this section, we introduce the proposed learning algorithm that is used to select each user's beamforming parameters. Our approach is based on using neural networks to learn the non-linear correlation between the channel characteristics and the beamforming parameters. In what follows, we separately discuss the several steps that are related to the design and application of our algorithm.

\subsection{GENERATION OF TRAINING/CSI DATA}
To construct the training data essential for the training phase, we generate random instances of the channels established between the BS and the two users, employing the statistical channel model delineated in \eqref{kanaliRayleigh}. Such an approach is also followed in \cite{zhao_power_2020,van_chien_power_2020,sanguinetti_deep_2018,sun_learning_2018,dandrea_uplink_2019,8935405,9134393} for the purposes of designing DL-based resource allocation and signal processing algorithms. This approach allows us to apply our DL-based methodology even in the absence of real channel data. Moreover, in practice, the Rayleigh fading channel model has been found to model wireless/mobile propagation channels accurately \cite{tse_fundamentals_2005}. The channels that are created with this procedure will be used to form the input of the neural network, as we will explain in the following parts of this section.

\subsection{SOLVING THE OPTIMAL BEAMFORMING PROBLEM FOR THE TRAINING DATA}

In order for our algorithm to learn how to infer beamforming decisions, the optimal beamformer decision for each one of the training CSI data is required. As a result, for each one of the training data instances, we need to solve the corresponding instance of problem \eqref{newargminmax}. However, as already explained, optimally solving problem \eqref{newargminmax} is not straightforward, mainly due to the fact that the convexity of the problem cannot be assumed. As a result, the application of any standard iterative algorithm for solving problem \eqref{newargminmax} is not guaranteed to provide the globally optimal solution. To overcome this problem, we solve problem \eqref{newargminmax} using a Constrained Optimization (CO) algorithm. More specifically, we use MATLAB's optimization toolbox \cite{Optimization_Toolbox} and the corresponding functions for constrained nonlinear multivariable problems like "\textbf{fmincon}". For each CSI instance, we apply the constrained optimization procedure (CO) 20 times for 20 random initial points, and we keep the response that minimizes $\Psi$ among the two users. This way, for each channel realization, we obtain the optimal beamforming parameters ($\rho_1,\rho_2,\delta_1,\delta_2, \tau_1,\phi_1,\phi_2 $). These parameters will be used as the network's outputs/labels. The proposed CO method cannot ensure that the global minimum will be found universally but is able to choose the best solution out of several local minima.

\begin{figure}[t]
\includegraphics[width=\linewidth]{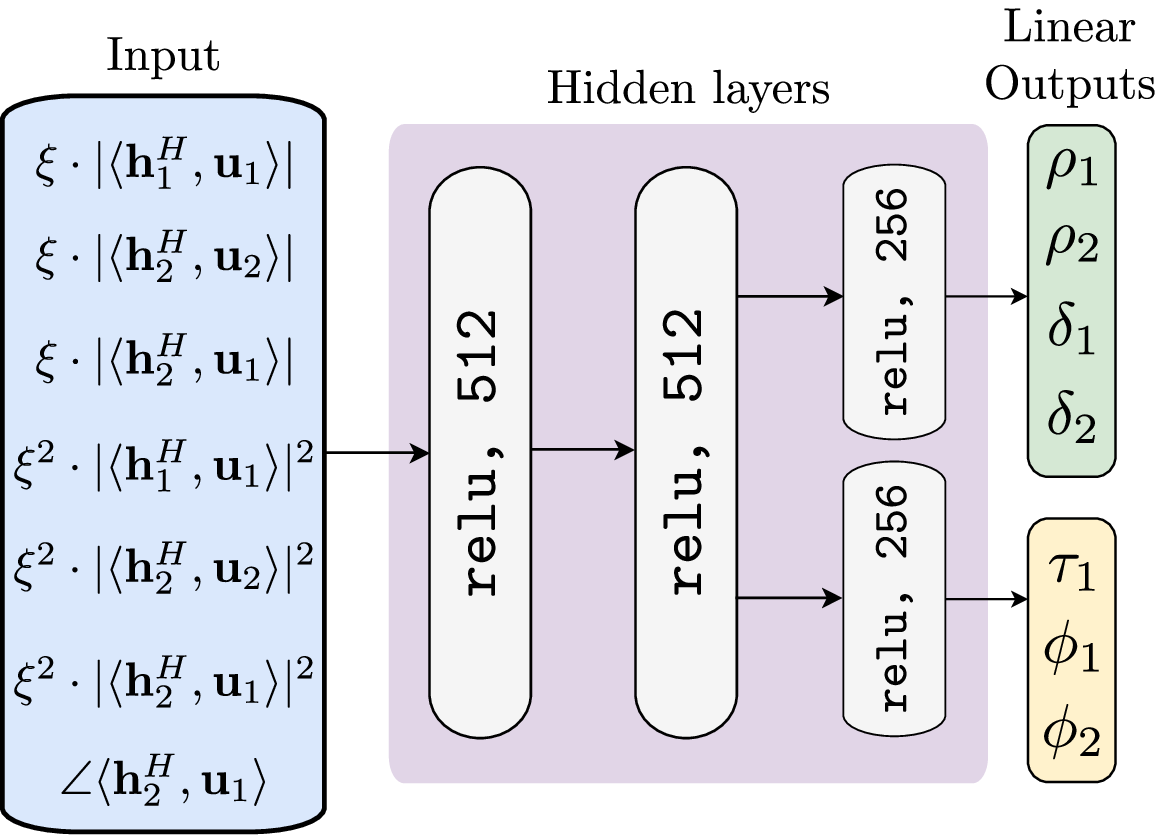}
\caption{Architecture of the used Neural Network. In this figure, we present the architecture of the proposed NN and give information about the shape of the input layers, the size and the activation functions of the hidden layers, and the linear outputs that are formed.}
\label{fig:NNpaper}
\end{figure}

\subsection{SELECTING APPROPRIATE NEURAL NETWORK ARCHITECTURES}

The literature contains multiple proposals for solving the power allocation problem using deep neural networks (DNN). In \cite{zhao_power_2020}, deep convolutional neural networks are used to perform min-max power allocation in a cell-free massive MIMO system. The use of convolutional layers is also proposed in \cite{van_chien_power_2020}, where a residual network consisting of multiple convolutional and fully connected layers is used for power allocation in a multi-cell massive MIMO setup. On the other hand, in \cite{loquercio_dronet_2018}, a Residual Network (ResNet) is used in order to predict angles similar to the ones in our problem.

Another neural network architecture commonly used in communications systems design, particularly in power allocation problems, is the fully connected one. In \cite{sun_learning_2018}, a fully connected DNN is introduced for the purposes of power allocation and interference management. The proposed scheme uses the magnitude of the channel realization as inputs of the DNN, achieving very good performance levels. Additionally, feed-forward neural networks with fully connected layers are employed in \cite{sanguinetti_deep_2018,dandrea_uplink_2019} and exploit user positions or shadowing coefficients as input to extract the optimal power allocation parameters.  

Given the above promising results of fully connected DNNs and the universal function approximation property \cite{hornik_multilayer_1989,Goodfellow-et-al-2016}, in our work, we propose the use of a fully connected DNN to solve problem \eqref{newargminmax}. The decision to use a fully connected DNN instead of other network architectures like Convolutional Neural Networks (CNN), Recurrent Neural Networks (RNN), or Long Short-Term Memory networks (LSTM) is that problem \eqref{newargminmax} has little structure to explore. In cases where specific characteristics emerge from a problem, using a CNN or an LSTM might prove beneficial. For example, in \cite{spatialdeep}, we can see that CNN is useful for problems that exploit the geographical location of the transmitter and the receiver to address a scheduling problem, and in \cite{nikasIlias}, we see that the use of an LSTM is providing better performance than a fully connected DNN in the case that the input data are consecutive channel matrices of a single user that is moving in high speed. 

Regarding the number of hidden layers, we set it equal to 3, and we determine the designated sizes of these layers in Fig. \ref{fig:NNpaper}. All hidden layers use the Rectified Linear Unit (ReLU) activation function. In order to decrease the network's complexity and processing time, the architecture is partly shared between the two tasks. However, the final hidden layer is divided into two smaller layers. Each of these smaller layers ends up in one of the two outputs, as we can observe in Fig. \ref{fig:NNpaper}. The first branch of the output includes the four parameters that take values in the range $[0,1]$ (i.e. $\rho_1, \rho_2, \delta_1, \delta_2$), and the second branch includes the three angles that take values in the range $[0,2\pi)$ (i.e. $ \tau_1, \phi_1, \phi_2 $). Although the prediction of all these outputs is a regression problem, due to their different nature and value range, we propose the separation of the last layer into two independent compartments. This approach has resulted in better performance for similar problems like those presented in \cite{loquercio_dronet_2018}. Both NN outputs use linear activation functions. Since the task that we are trying to accomplish is a regression problem, the mean absolute error (MAE) loss function is preferred by both outputs. Subsequently, the overall loss function of the neural network is a weighted sum of the two individual losses. In our case, the individual losses contribute equally to the overall loss of the network since we use the sum of the individual loss values as the overall one. We train the neural network using the ADAM algorithm with Nesterov's momentum \cite{sutskever_importance_2013,dozat_incorporating_2016}. The training continues for a maximum of 200 epochs unless it meets the predetermined training conditions. In this latter case, it stops automatically. 

Although we suggest a specific DNN structure, it is essential to note that discovering the optimal neural network architecture can be viewed as a separate optimization problem that requires additional research. Based on the approximation and estimation bounds for NNs presented in \cite{barron_approximation_1994} and the works of \cite{Goodfellow-et-al-2016} in deep learning techniques, we experimented with various architectures of fully connected networks, aiming to find the best one in terms of MAE. The range of the fully connected NNs we tested varies from two to six layers and consists of different numbers of neurons. During our tests, we also used various activation functions for the hidden layers, as well as for the output layer and different training algorithms. We also examined using two separate neural networks, one for each output, and compared these results with our proposed layout.

\subsubsection{Input data}
As the input of the DNN, we use the magnitude and the angles of the dot products between the channel realizations of the two NOMA users and the vectors of the orthonormal basis, as presented in Fig. \ref{fig:NNpaper}. That is, as inputs will be used the terms $ | \langle  \mathbf{h}_1^H,\mathbf{u}_1 \rangle | \;,| \langle  \mathbf{h}_2^H,\mathbf{u}_2 \rangle | \;, | \langle  \mathbf{h}_2^H,\mathbf{u}_1 \rangle |$ and their squared values, as well as the term $\angle\langle  \mathbf{h}_2^H,\mathbf{u}_1 \rangle$. We note that we do not use the angles of $\langle  \mathbf{h}_1^H,\mathbf{u}_1 \rangle$ and  $\langle  \mathbf{h}_2^H,\mathbf{u}_2 \rangle$ for the input of the NN, because they are equal to zero, due to the properties of the orthonormal basis. We use $\xi$ as an equalization factor so that all the input components belong to similar orders of magnitude, as shown in Fig. \ref{fig:NNpaper}. In our case, we set $\xi$ to the value $10^6$. Based on this pre-processing procedure, the neural network will have the same input size for every different antenna scheme, with dimensions $7 \times 1$. This property allows us to train a single neural network for all the different antenna schemes that we consider. It also enables the opportunity to use the same neural network in case a new antenna scheme appears just by merging the data from all the antenna schemes and retraining the same neural network.

\subsubsection{Handling of the testing stage outputs}

As we mentioned above, the values that the beamforming parameters can take are limited inside specific ranges. Namely, parameters $\rho_1, \rho_2, \delta_1$ and $ \delta_2$ can take values in the range $[0,1]$ and parameters $ \tau_1, \phi_1 $ and $  \phi_2 $ in the range $[0,2\pi)$. Additionally, the values of the parameters have to satisfy constraints \eqref{constr1}, \eqref{constr2}, and \eqref{constr3}. To ensure this is the case, we check every network output at the test stage. If any of the constraints is not satisfied and/or the values are not in the designated ranges, we handle the values of the parameters to make the solution feasible. All the steps of this procedure are illustrated in Algorithm \ref{alg1}. We note that the values of variables $\hat{\rho}_{i},\hat{\delta}_{j}$ and $\kappa_i$ of Algorithm \ref{alg1} are set during the simulation part of our work. During this algorithm, we initially check if the values of $\rho_1, \rho_2, \delta_1$ and $ \delta_2$ exceed the range limits and if they satisfy \eqref{constr3}. Subsequently, we check if the expressions of \eqref{constr1} and \eqref{constr2} are defined and if the constraints are met. Finally, we check again if \eqref{constr3} is satisfied after the adjustments and return the new output values. Using this post-neural-network algorithm, we ensure all the constraints and ranges are met. After testing tens of thousands of inputs and outputs, we have to mention that we never encountered a case where the predicted parameters did not meet the constraints or limits. This means that the network can learn and predict the parameters of the beamforming vectors for this system model well. Nevertheless, this may not be the case for system models with different parameters; thus, we introduce Algorithm \ref{alg1}.

\begin{algorithm}
\caption{Post-Neural-Network Algorithm}
\begin{algorithmic}[1]
\label{alg1}
\REQUIRE $\rho_1, \rho_2, \delta_1, \delta_2, \tau_1, \phi_1, \phi_2, \mathbf{h}_2, \mathbf{u}_1, \mathbf{u}_2, \hat{\rho}_{1}, \hat{\rho}_{2}, \hat{\delta}_{1}, \hat{\delta}_{2},$ \\ $ \; \; \; \; \; \; \; \; \;  \kappa_1, \kappa_2$ 

\FOR{$i \gets 1$ \TO $2$} 
    \IF{$\rho_i \leq 0$}
        \STATE $\rho_i \gets \hat{\rho}_{i}$
    \ENDIF
\ENDFOR

\FOR{$j \gets 1$ \TO $2$}
    \IF{$\delta_j \leq 0$}
        \STATE $\delta_j \gets \hat{\delta}_{j}$
    \ENDIF
\ENDFOR

\STATE $S \gets \rho_1^2 + \rho_2^2 + \delta_1^2 + \delta_2^2$ 
\IF{$S > 1$}
    \STATE $\rho_1 \gets \frac{\rho_1}{\sqrt{S}}$
    \STATE $\rho_2 \gets \frac{\rho_2}{\sqrt{S}}$
    \STATE $\delta_1 \gets \frac{\delta_1}{\sqrt{S}}$
    \STATE $\delta_2 \gets \frac{\delta_2}{\sqrt{S}}$
\ENDIF

\IF{$\cos(\tau_1) - \sin(\tau_1) = 0$}
    \STATE $\tau_1 \gets \tau_1 + \kappa_1$
\ENDIF

\WHILE{\eqref{constr1} is not met}
    \STATE $\delta_1 \gets \frac{\delta_1}{\rho_1 + 1}$
\ENDWHILE

\STATE $z \gets |\delta_1 e^{j\phi_1} \mathbf{h}_2^H \mathbf{u}_1 + \delta_2 e^{j\phi_2} \mathbf{h}_2^H \mathbf{u}_2|$

\IF{$z = 0$}
    \STATE $\delta_2 \gets \delta_2 + \kappa_2$
\ENDIF

\WHILE{\eqref{constr2} is not met}
    \STATE $\delta_1 \gets \frac{\delta_1}{\rho_1 + \rho_2 + 1}$
    \STATE $\delta_2 \gets \frac{\delta_2}{\rho_1 + \rho_2 + 1}$
\ENDWHILE

\STATE $S \gets \rho_1^2 + \rho_2^2 + \delta_1^2 + \delta_2^2$ 

\IF{$S > 1$}
    \STATE $\rho_1 \gets \frac{\rho_1}{\sqrt{S}}$
    \STATE $\rho_2 \gets \frac{\rho_2}{\sqrt{S}}$
    \STATE $\delta_1 \gets \frac{\delta_1}{\sqrt{S}}$
    \STATE $\delta_2 \gets \frac{\delta_2}{\sqrt{S}}$
\ENDIF

\RETURN $\rho_1, \rho_2, \delta_1, \delta_2, \tau_1$
\end{algorithmic}
\end{algorithm}

\section{SIMULATION \& NUMERICAL RESULTS}
In order to evaluate the performance of the proposed learning algorithms, we considered a scenario where a multi-antenna BS communicates with two users following the channel model presented in \eqref{kanaliRayleigh}. Note that the specific channel model is part of 3GPP standard channel models \cite{3gpp.36.942}. As a result, it can be considered a good option for generating channel instances that preserve some of the properties of real channel data measurements. The remaining parameters of our simulation-driven performance evaluation are presented in Table \ref{tab:table2}. Concerning user placement, for our purposes, we have considered the case that $U_1$ is always located at a distance between 600 and 650 meters from the BS at each realization and $U_2$ at a distance between 350 and 400 meters. Moreover, we have assumed that the modulation order of $U_1$ and $U_2$ is the same and is equal to $M = M_1 = M_2 = 4$, while the noise variance was also assumed to be the same for both users and equal to $N_{PD} \times B$, where $N_{PD}$ the noise density and $B$ the system bandwidth, both given in Table \ref{tab:table2}. For this particular simulation scenario, we have investigated the performance of our proposed algorithm and compared it with that of commonly used beamformers, which serve as benchmarks for our algorithm. These beamformers are briefly presented in what follows.

\begin{table}[h!]
  \centering
\caption{Model parameters used in simulations\label{tab:table2}}
\centering
\begin{tabular}{|c|c|}
\hline
Parameter & Value\\
\hline
Carrier frequency & 2GHz\\
\hline
System Bandwidth ($B$) & 10MHz\\
\hline
Noise power density ($N_{PD}$)& -174 dBm/Hz\\
\hline
Path-loss in dB & PL = 128.1 + 37.6$\log_{10}$d, d in km\\
\hline
Transmit power & 100mW\\
\hline
$U_1$ distance & 600m - 650m\\
\hline
$U_2$ distance & 350m - 400m\\
\hline
$M_1, M_2$ & 4\\
\hline
\end{tabular}
\end{table}

\subsection{THE CONSIDERED BENCHMARKS}
The beamforming algorithms which were compared, in terms of performance, to our proposed algorithm were the following.

\subsubsection{Maximum Ratio Transmission (MRT)}
In the case of MRT \cite{lo_maximum_1999}, the beamforming vector applied on the data of user $U_k$ is expressed as:
\begin{equation}
\mathbf{w}_k = \frac{\mathbf{h}_k}{|| \mathbf{h}_k||}.
\end{equation}
MRT beamforming aims at maximizing the Signal to Noise Ratio (SNR) of the signal $s_k$ at user $U_k$. The main disadvantage of this approach is that it does not take into account the interference caused by the weak user, which can be large.

\subsubsection{Zero Forcing Beamforming (ZFBF)} 
ZFBF aims to create beamforming vectors that cause zero interference to other users \cite{bjornson_optimal_2013}. To do so, the beamforming vector used for transmitting to the $n$-th user is orthogonal to the channel vectors of all other users. In our case, this implies that the following is true:
\begin{equation}
\mathbf{h}_1^H\mathbf{w}_2 = 0 \;\;,\;\; \mathbf{h}_2^H\mathbf{w}_1 = 0 .
\end{equation}

 While one can argue that ZFBF completely eliminates the interference of each user in the receiving signal of the other one and, as a result, is not exploiting the concepts of non-orthogonal access, our simulation results indicate that ZFBF can still result in reliable performance while avoiding the computational complexity of constraint optimization techniques. ZFBF is based on the beamformers' orthogonal relationship with the other users' channel vectors. This property is enhanced as the number of transmitting antennas is increasing compared to the number of users since this increase offers more degrees of freedom at the beamforming vector. However, in multiple antenna NOMA systems, many user pairs or clusters coexist. In these cases, ZFBF is not used as a beamforming technique for each user of each cluster. Usually, it is used as the first step of the beamforming design in order to reduce inter-cluster interference. But again, in most of these cases, the number of users and receive antennas surpasses the number of transmit antennas. Thus, the zero-interference properties of ZFBF cannot be achieved even with ZFBF that tackles inter-cluster interference, and therefore, this technique is considered sub-optimal \cite{ali_non-orthogonal_2017,choi_minimum_2015}.

\subsection{NUMERICAL RESULTS}
In this section, we compare the performance of our proposed DL model (NN) with the performance of CO and the benchmarks. Each technique is tested for four antenna schemes (2,3,4 and 5 antennas at the BS). For the training of the NN, we used fifty thousand realizations for each antenna scheme, creating a dataset of two hundred thousand input/output pairs in total. The two benchmark techniques were used as follows. First, we used MRT for both users (MRT), then the ZFBF again for both users (ZFBF), and finally, we used combinations of these two choices between the users. That is, MRT for the first user and ZFBF for the second one (MRT 1 \& ZFBF 2) and vice versa (ZFBF 1 \& MRT 2). For the techniques MRT, MRT 1 \& ZFBF 2, and ZFBF 1 \& MRT 2, we do not have the theoretical expression to calculate the BER, so we calculated the maximum BER for each technique with the use of simulations. Specifically, for each channel realization that we tested these three techniques, we used forty million symbols in order to extract the value of the maximum BER.

Subsequently, we set the values of the variables $\hat{\rho}_{i},\hat{\delta}_{j}$ and $\kappa_i$ for the post-neural-network Algorithm \ref{alg1} that ensures that all the problem constraints and ranges are met. The values of $\hat{\rho}_{i}$ and $\hat{\delta}_{j}$ are chosen to be the mean values of the corresponding $\rho_{i} $ and $\delta_{j}$ parameters from the training dataset of each antenna scheme. We note that although the initialization of this algorithm can be done with any set of values that follows the range constraints for $\rho_{i} $ and $\delta_{j}$, we propose these specific ones due to the narrow distribution that each of the variables $\rho_{i} $ and $\delta_{j}$ follows in this particular training dataset. Finally, we have $\kappa_1 = 10^{-5}$ and $\kappa_2 = 10^{-5}$. The values of $\kappa_i$ are chosen to be relatively small since these parameters are only used to ensure that the denominators of the expressions \eqref{constr1} and \eqref{constr2} are non-zero.  

\begin{figure}[h!]
\includegraphics[width=\linewidth]{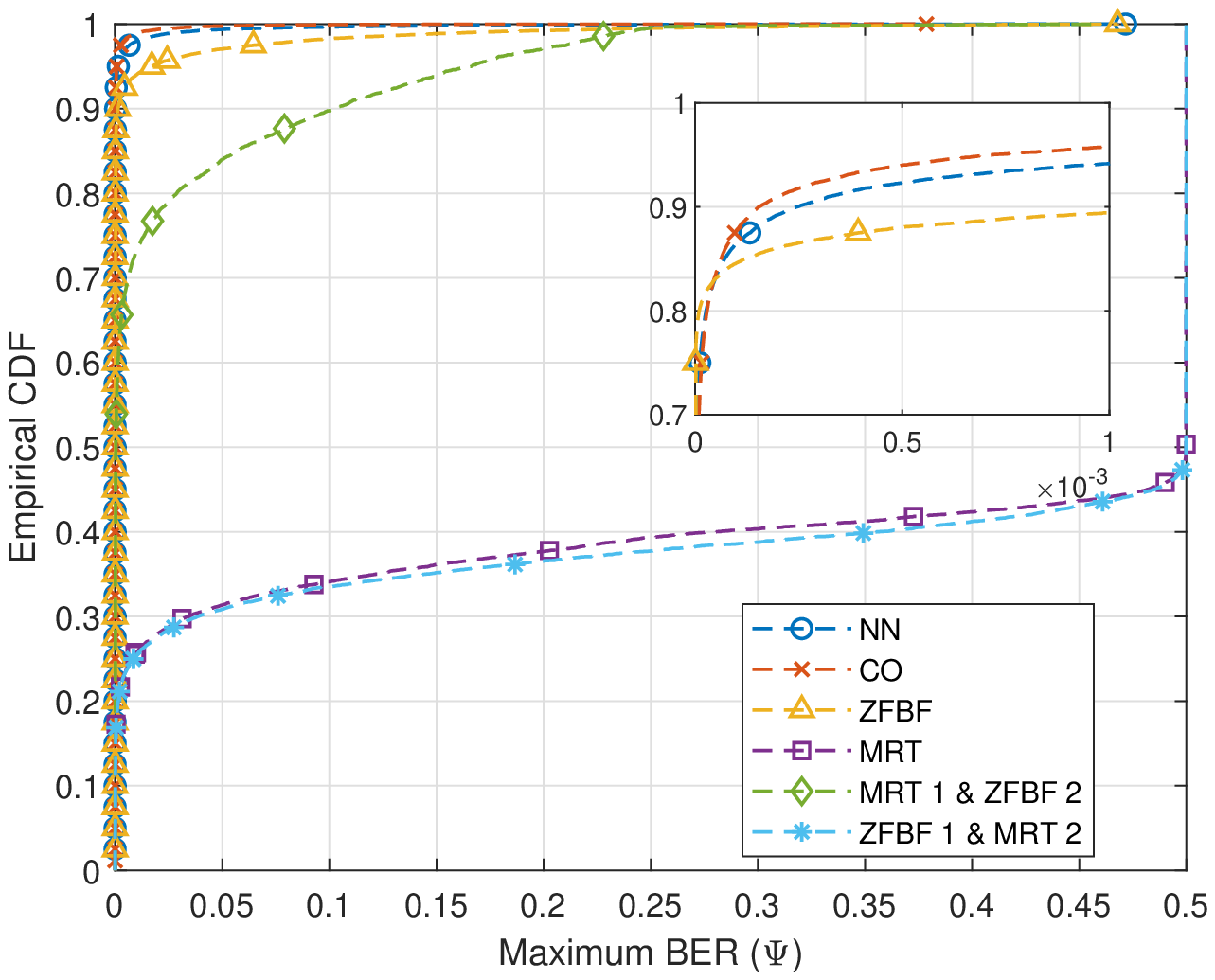}
\caption{Empirical CDF comparison of $\Psi$ for the two users for various beamforming techniques (the smaller figure is a zoomed version of the original figure that focuses on specific $\Psi$ and percentage values).}
\label{fig:BERALL}
\end{figure}

\begin{table*}[b!]
\centering
\caption{ $\Psi$ values of every beamforming technique, for designated percentage values of the ECDF plot in  Fig. \ref{fig:BERALL}.\label{tab:table3}}
\centering
\begin{tabular}{|c||c|c|c|c|c|c| }
\hline
\diagbox{Percentage }{Algorithm}Percentage & NN & CO & ZFBF & MRT & MRT 1 \& ZFBF 2 & ZFBF 1 \& MRT 2 \\
\hline
85 \% &  $7.1 \times 10^{-5}$ & $6.2 \times 10^{-5}$ & $1.2  \times 10^{-4}$ & $\approx 0.5$ &   0.056 & $\approx 0.5$\\
\hline
90 \% & $2 \times 10^{-4}$ & $1.5 \times 10^{-4}$ & $ 1.3 \times 10^{-3}$ & $\approx 0.5$ & $ 10^{-1}$ & $\approx 0.5$ \\
\hline
95 \% & $10^{-3}$ &  $7 \times 10^{-4}$ & $1.7 \times 10^{-2}$ & $\approx 0.5$ & 0.16  &$\approx 0.5$ \\
\hline
99 \% & 0.03 & 0.01 & 0.16 & $\approx 0.5$ & 0.23 & $\approx 0.5$\\
\hline
\end{tabular}
\end{table*}

In Fig. \ref{fig:BERALL}, we present an Empirical Cumulative Distribution Function (ECDF) plot with the values of the maximum BER of the two users for all these techniques for all the different antenna schemes. The results that were used in order to generate these ECDF plots consider five thousand randomly positioned user pairs for each antenna scheme, creating a total of twenty thousand different realizations. By merging the data of all the different antenna schemes in each case, we present a catholic comparison of the beamforming techniques.  We need to note that the BER values of MRT, MRT 1 \& ZFBF 2, and ZFBF 1 \& MRT 2 techniques vary from 0.5  down to $10^{-7}$ because the results from these techniques come from simulations since we do not have a theoretical expression. Thus, every BER value under this threshold is going to be equal to zero for these techniques.

\begin{figure}[h]
\includegraphics[width=\linewidth]{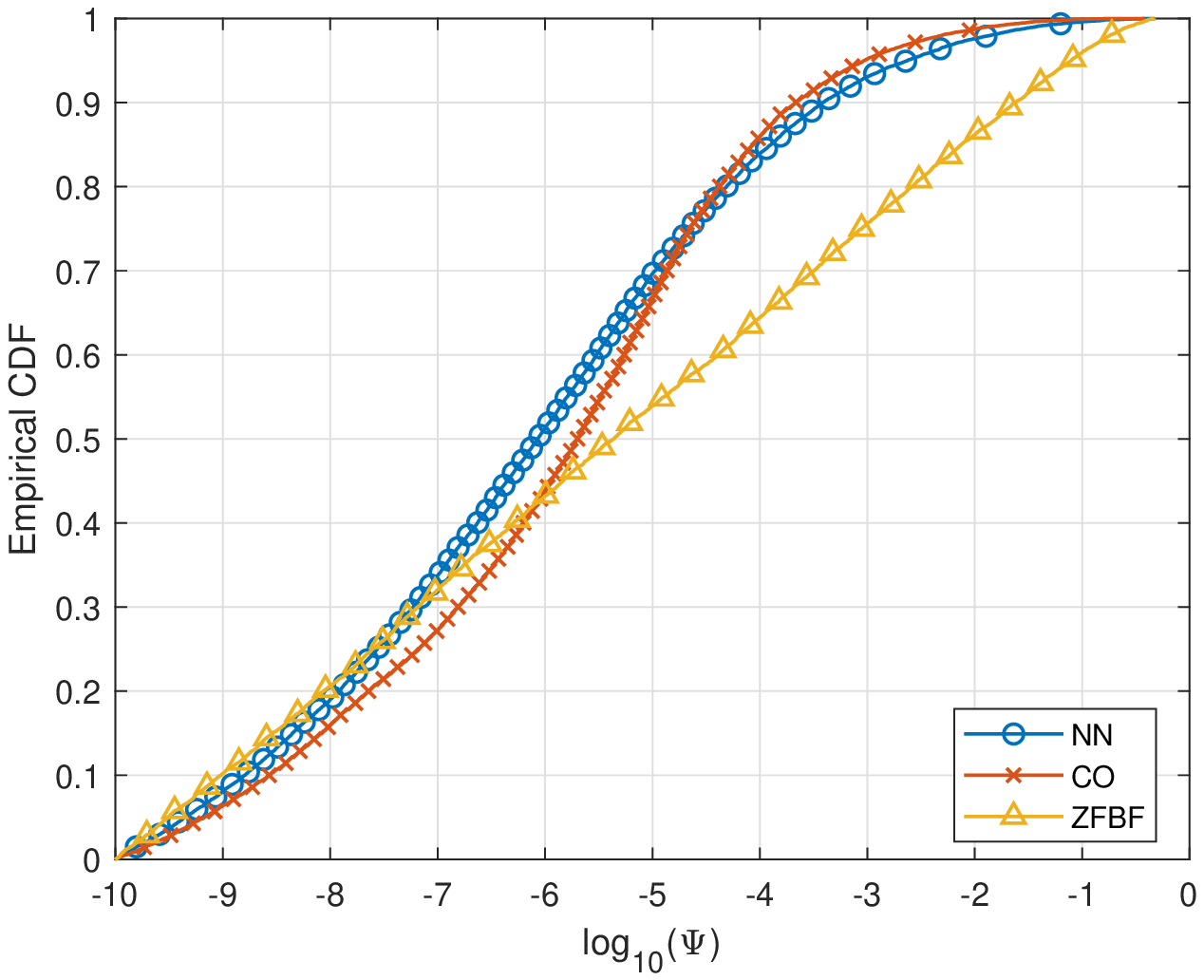}
\caption{Empirical CDF of the logarithmic values of $\Psi$ for the two users for different beamforming techniques (BER values from $10^{-10}$ to $0.5$).}
\label{fig:hist2}
\end{figure}

Initially, we have to notice in Fig. \ref{fig:BERALL} the superiority of CO, NN, and ZFBF against the remaining three techniques. We observe that these first three techniques have a vast majority of their cases concentrated in low $\Psi$ (lower than 0.05). On the other hand, the other three techniques (i.e. MRT for both users, MRT for the first user and ZFBF for the second one and ZFBF for the first user and MRT for the second user) have a large number of cases with low $\Psi$, but also have a big portion of realizations that result to BER of the highest value. A fact that dramatically affects their overall performance, as we can also see in Table \ref{tab:table3} that presents the values of the ECDF of each beamforming technique. A more accurate image for the performance of the three best-behaving techniques (i.e. NN, CO, ZFBF) can be seen in the zoomed box within the ECDF plot of Fig. \ref{fig:BERALL}. There, we observe that the NN technique results in the best results from the other techniques, only second behind the constrained optimization (CO) algorithm. In addition, in Fig. \ref{fig:hist2}, we present the ECDF plot for the three techniques for which we can theoretically calculate their $\Psi$ values. As we can see, NN continues to be the best-performing technique alongside CO for log$_{10}(\Psi)$ values down to -10 (i.e. $\Psi$ equal to $10^{-10}$). Especially for some log$_{10}(\Psi)$ values, we can observe that NN has a distribution even better than CO. On the other hand, ZFBF seems to have similar results with the other two techniques for log$_{10}(\Psi) < -8$ values but has a significant gap in performance for log$_{10}(\Psi) > -6$ values.

As expected, CO can find suitable beamforming vectors by computing the parameters of the beamforming structure that we present in \eqref{orth_basis} and \eqref{eq:w_1_w_2}. With the help of the neural network that we proposed, these parameters can be learned and used to create beamforming vectors in any MISO-NOMA system that is based on user fairness. However, we need to emphasize the time that these last two techniques (i.e. CO, NN) need in order to provide these results. The calculation of the optimal beamformer, in our case, is the task of constrained optimization with multiple random initialization points in a non-convex seven-dimensional space. In combination with the fact that the BER expressions consist of multiple sums, of which the derivatives have to be calculated, we understand that finding the optimal solution through constrained optimization is a procedure that requires a lot of computational time. Compared to CO, the pre-trained NN with size such as in our case, is a robust procedure with low complexity and needs substantially less time in order to give a solution.

\begin{figure}[h!]
\includegraphics[width=\linewidth]{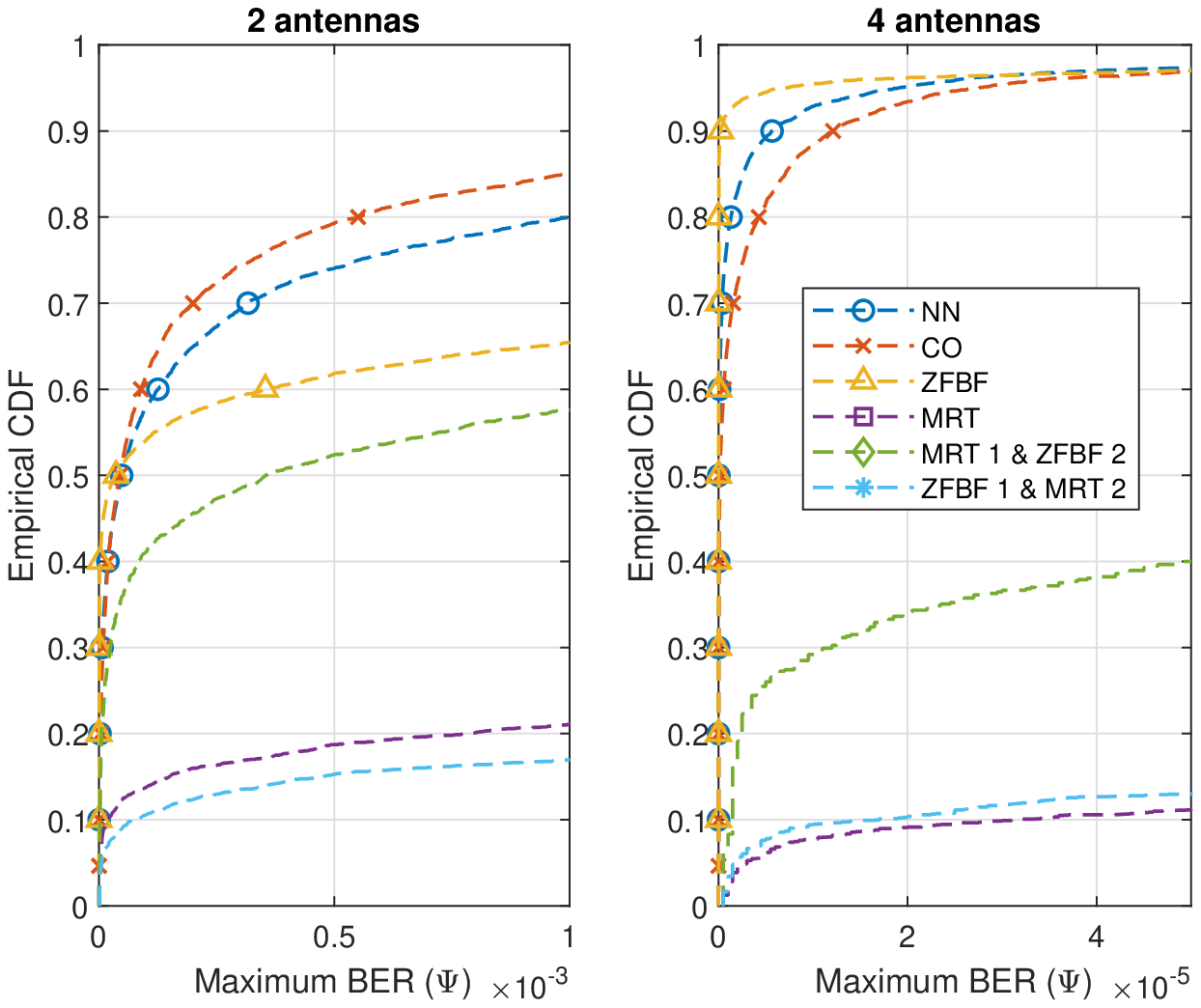}
\caption{Comparison of $\Psi$ for the two users for various beamforming techniques into two specific antenna schemes. Left: 2-antenna scheme. Right: 4-antenna scheme.}
\label{fig:BERALL7}
\end{figure}

In Fig. \ref{fig:BERALL7}, we see the ECDF plots for the BER values in two of the antenna schemes that we use. First, we see that in the 2-antenna scheme, the NN is again second best after CO. In this case, we see a more considerable difference in performance from the following best technique than in the previous figures. As we can see, over 80\% of the NN predictions result in BER lower than $10^{-3}$, in contrast to almost 65\% of the examples of ZFBF. On the other hand, in the case of the 4-antenna scheme, we see that ZFBF is slightly better. This was expected due to the nature of ZFBF. As we noticed before, ZFBF is not a NOMA technique since that uses SIC to eliminate the interference, but tries to avoid it altogether. Thus, by increasing the number of antennas, we offer more degrees of freedom to this technique in order to find a vector orthogonal to the other user's channel. However, we can observe that although ZFBF has a very good performance, NN comes second close behind it, with a negligible BER difference between them that is less than $0.5 \times 10^{-5}$ at 90\% of the cases. In this case, we also see that NN is behaving better than CO. This can be a random event due to differences between the predictions of the NN and CO that result in better NN behavior in this specific range of values.

\begin{figure}[h!]
\includegraphics[width=\linewidth]{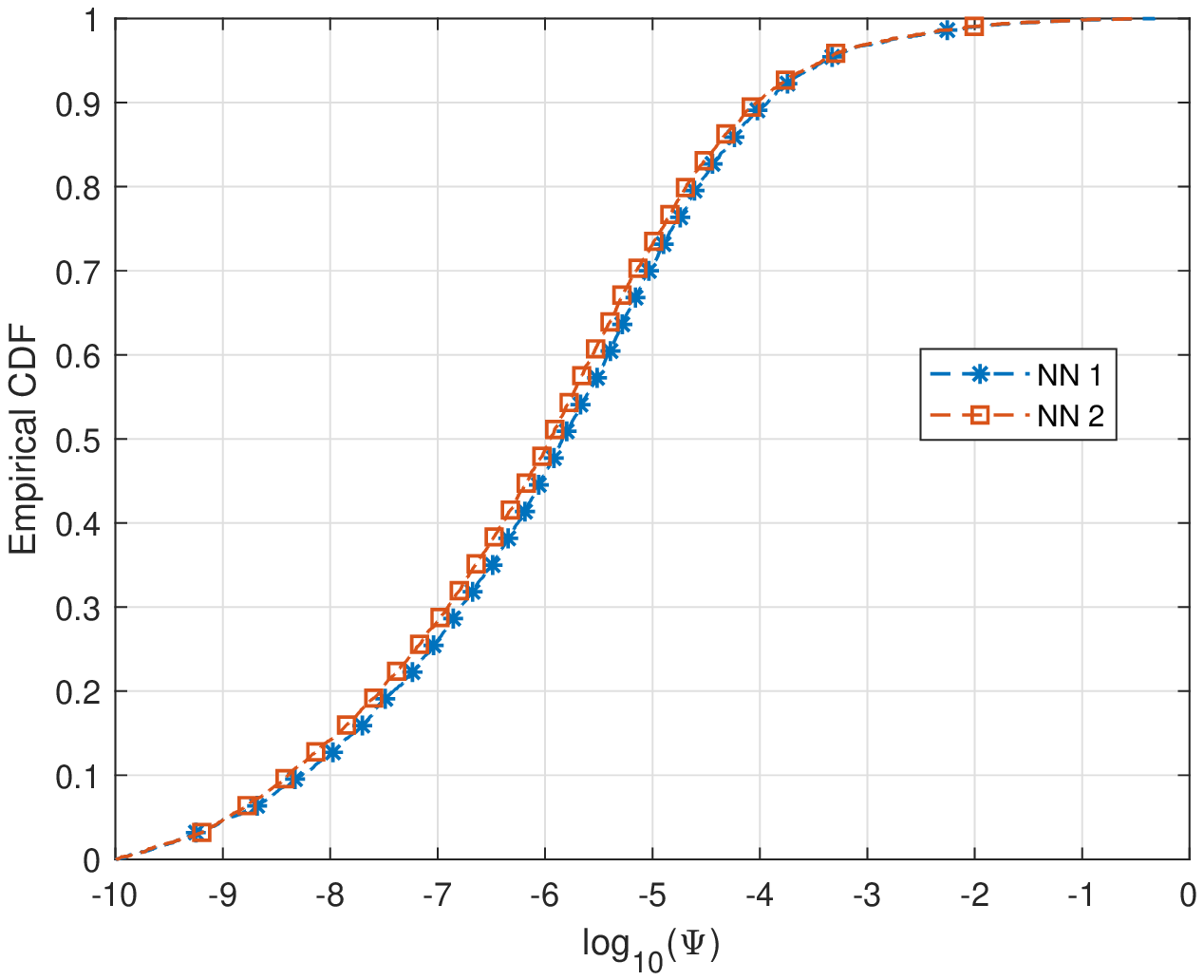}
\caption{ Empirical CDF plot for the performance comparison of the 3-antenna scheme between NN 1 that is trained with the data of all the available antenna schemes (2,3,4 and 5 antennas) and a NN 2 that is trained only with data from the 3-antenna scheme.}
\label{fig:hist3}
\end{figure}

Finally, in our algorithm, we use as the input of the NN the inner products between the channel vectors and the orthonormal basis vectors. This means that the size of the input for the NN will remain the same for any antenna scheme and that we can train a single NN with data from various antenna schemes. An example of this property we present in the ECDF plot of Fig. \ref{fig:hist3} where we compare the logarithm of the performance of different channel pairs of the 3-antenna scheme. In Fig. \ref{fig:hist3}, we can see that the use of the NN 1 that is trained with data from all the available antenna schemes (2,3,4 and 5 antennas) does not affect at all the $\Psi$ performance of the system when compared with NN 2 that is explicitly trained with data from the 3-antenna scheme. If anything, it performs equally well. Thus, with our algorithm, we can train one single NN with data from multiple antenna schemes and use it for all these antenna schemes with great results. A technique that is both robust and efficient.

 \begin{figure}[h!]
\includegraphics[width=\linewidth]{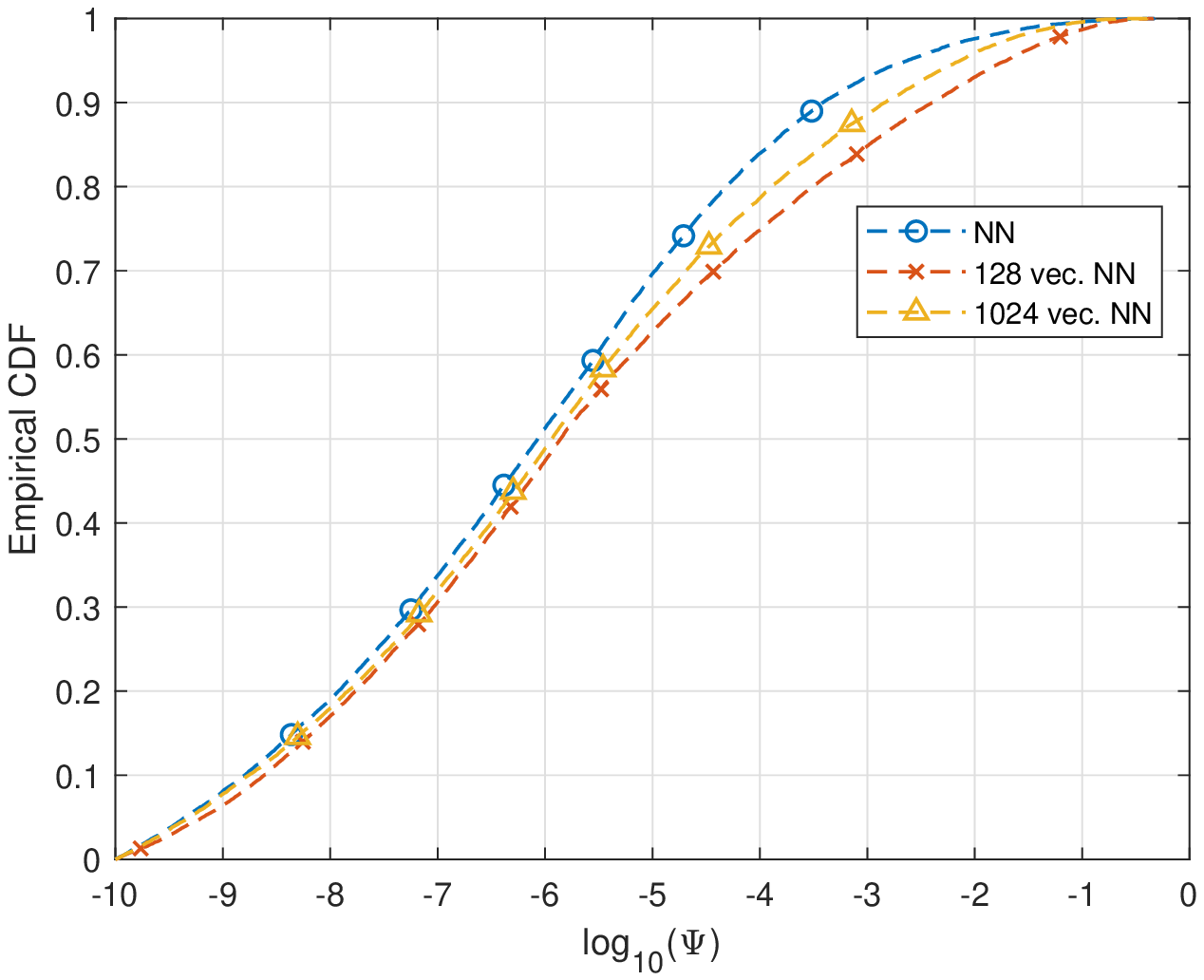}
\caption{ Empirical CDF plot for the performance comparison a NN that is trained \& tested with the produced data, a NN that is trained \& tested with 128 candidate vectors that are produced with \textit{k-means} from the data, a NN that is trained \& tested with 1024 candidate vectors that are produced with \textit{k-means} from the data.}
\label{fig:RVQ}
\end{figure}

Another element of robustness and efficiency of our system can be seen in Fig. \ref{fig:RVQ}. In this case, we compare three different NNs. First is the initial NN that we used for all of our simulations. In the case of the other two networks, the input data are not the vectors that were created through the pre-processing procedure of the exact channel vectors. For these networks, the procedure of input creation is as follows. First, we use the pre-processing procedure on the exact channel vectors and create the input vectors. Second, we run a \textit{k-means} clustering algorithm on the input vectors that correspond to the training data. Then, we replace each input vector with the centroid vector of the cluster to which it belongs and train the network with these \textit{k} representative centroid vectors. For the testing process, we replace each testing input vector with the closest centroid vector in terms of Euclidean distance. The \textit{k} value of the \textit{k-means} algorithm that defines the number of the representative centroid vectors is equal to 128 in the case of the second network and 1024 in the case of the third one. We note that the output/label data of the training procedure have not been changed in any of these cases. That way, during the training process of the second and third networks, the model receives different labels for the same input. Therefore, during the testing phase of the networks, the outputs are going to be also representative vectors of each cluster. Finally, based on these outputs, we calculate the maximum BER with the use of the original channel vectors. In Fig. \ref{fig:RVQ}, it can be seen that the networks that use candidate vectors perform closely to the ones that exploit perfect channel knowledge, even though the predicted beamforming parameters do not correspond to the exact channel vector. Therefore, these models can be really useful in cases where the available training data are too noisy or have multiple minor variations.

\subsection{COMPUTATIONAL TIME COMPARISON}

As mentioned in previous sections, there is a significant difference between the computational time that CO needs in order to provide a prediction of the beamforming parameters and the computational time that NN needs. We run the CO optimization in MATLAB version 2022a, using the provided Optimization Toolbox \cite{Optimization_Toolbox}. We trained and tested the NN model in Python version 3.9.13 \cite{van1995python}, with the use of TensorFlow version 2.9.1 \cite{tensorflow2015-whitepaper} and Keras version 2.9.0 \cite{chollet2015keras}. Both of these algorithms run in a system with an Intel Core i7-10610U CPU and 16 GB of available RAM memory. In Table \ref{tab:time}, we present a comparison of the time consumption of the two algorithms. As can be seen, the proposed NN model is much faster than the CO algorithm by many orders of magnitude. Additionally, we see that the time that CO needs in order to predict the beamforming vector parameters is increasing proportionally to the increase of transmitting antennas. On the other hand, NN needs the same time for the prediction in any antenna scheme since the same NN is used for all of them.   

\begin{table}[h]
\centering
\caption{Average time consumption of two algorithms \label{tab:time}}
\begin{tabular}{|c|c|c|}
\hline
\diagbox{Number \\ of antennas}{Algorithm} & NN & CO \\ \hline
2 & 60 $\mu$s & 2.35 s \\ \hline
3 & 60 $\mu$s & 2.43 s \\ \hline
4 & 60 $\mu$s & 2.66 s \\ \hline
5 & 60 $\mu$s & 2.82 s \\ \hline
\end{tabular}
\label{table:diagonal}
\end{table}

\section{CONCLUSION}

In this paper, the closed-form expression of the BER for two users was expanded for the MISO-NOMA scenario over Rayleigh channels. A fairness-based optimization problem was introduced with the help of the derived BER expressions and the novel beamforming scheme that we proposed. Additionally, we suggested a DL model that learns the aforementioned optimization problem and predicts the beamforming parameters of the beamforming vectors. The structure of the input of our DL system makes it work successfully when different antenna schemes are used from the transmitter and can be easy to scale for other use cases. This quality reduces the cost of training and deploying a system like this, as using a different NN pair for each case is unnecessary. We present the performance of our system in terms of both maximum BER and computational time. Finally, we show the robustness of our learning system by testing its generalization capabilities and resilience to imperfections in the channel state information. 

Some extensions we would like to explore in future works are scenarios with a greater number of users and scenarios using real-time data. Additionally, a logical next step of our work can be the extension towards MIMO-NOMA systems. An extension like this would be based on a different beamforming scheme; therefore, the new BER expressions and the optimization problem could differ from those we provide. Thus, this analysis may not be necessarily straightforward to obtain.

\appendices

\section{Notations}
\label{section:Notations}

In this appendix, we offer Table \ref{tab:notations}, which summarizes all the significant notations of our work. 

\begin{table}[h!]
\centering
\caption{ Table for notations \label{tab:notations}}
\centering
\begin{tabular}{|c|c| }
\hline
Notation & Definition\\ 
\hline
$N_t$ & Number of antennas\\
\hline
$U_n$ & User $n$ \\ 
\hline
$s_n$ & QAM symbol of $U_n$ \\ 
\hline
$M_n$ & Modulation order of $U_n$ \\ 
\hline
$y_n$ & Received signal at $U_n$ \\
\hline
$\mathbf{x}$ & Transmitted signal \\ 
\hline
$\mathbf{h}_n$ & Channel vector between BS and $U_n$ \\
\hline
$\mathbf{w}_n$ & Complex beamforming vector for $U_n$ \\ 
\hline
$\nu_n$ & AWGN term at $U_n$ \\ 
\hline
$N_0$ & Noise variance \\ 
\hline
$\theta_{i,j} $  &  $ \angle \mathbf{h}_i^H\mathbf{w}_j$ \\
\hline
$\mathbf{u_i}$ & Orthonormal basis vectors \\
\hline
$\rho_i$, $\delta_i$ & Beamformer parameters $\in \mathbb{R}^{+} $\\
\hline
$\theta_i$, $\phi_i$ & Beamformer parameters $\in [ 0 , 2\pi) $ \\
\hline
$\tau_1$ & $\phi_1-\theta_1$ \\ 
\hline
$P_{e}^{(n)}$ & Average conditional BER at $U_n$\\
\hline
$P_{b,i}^{(1)}$ & BER for the $i$-th bit of a QAM symbol at $U_n$ \\
\hline
$\Psi$ & $\max\{P_{e}^{(1)} ,P_{e}^{(2)} \}$ \\
\hline
$Y_1 $ & $(\sqrt{M_2}-1)^2 \cdot (M_1 -1) / (M_2 - 1)  $\\
\hline
\end{tabular}
\end{table}

\section{Example $M_1 = M_2 = 4 $}
\label{section:Appendix}
In this appendix, we present the expressions of the conditional BER for the two users in the case of multiple antennas at BS (MISO - NOMA). We note that both users use a  4-QAM modulation, i.e., $M_1 = M_2 = 4$.

\subsection{$U_1$ EXPRESSION}
\begin{figure}[h!]
\includegraphics[width=\linewidth]{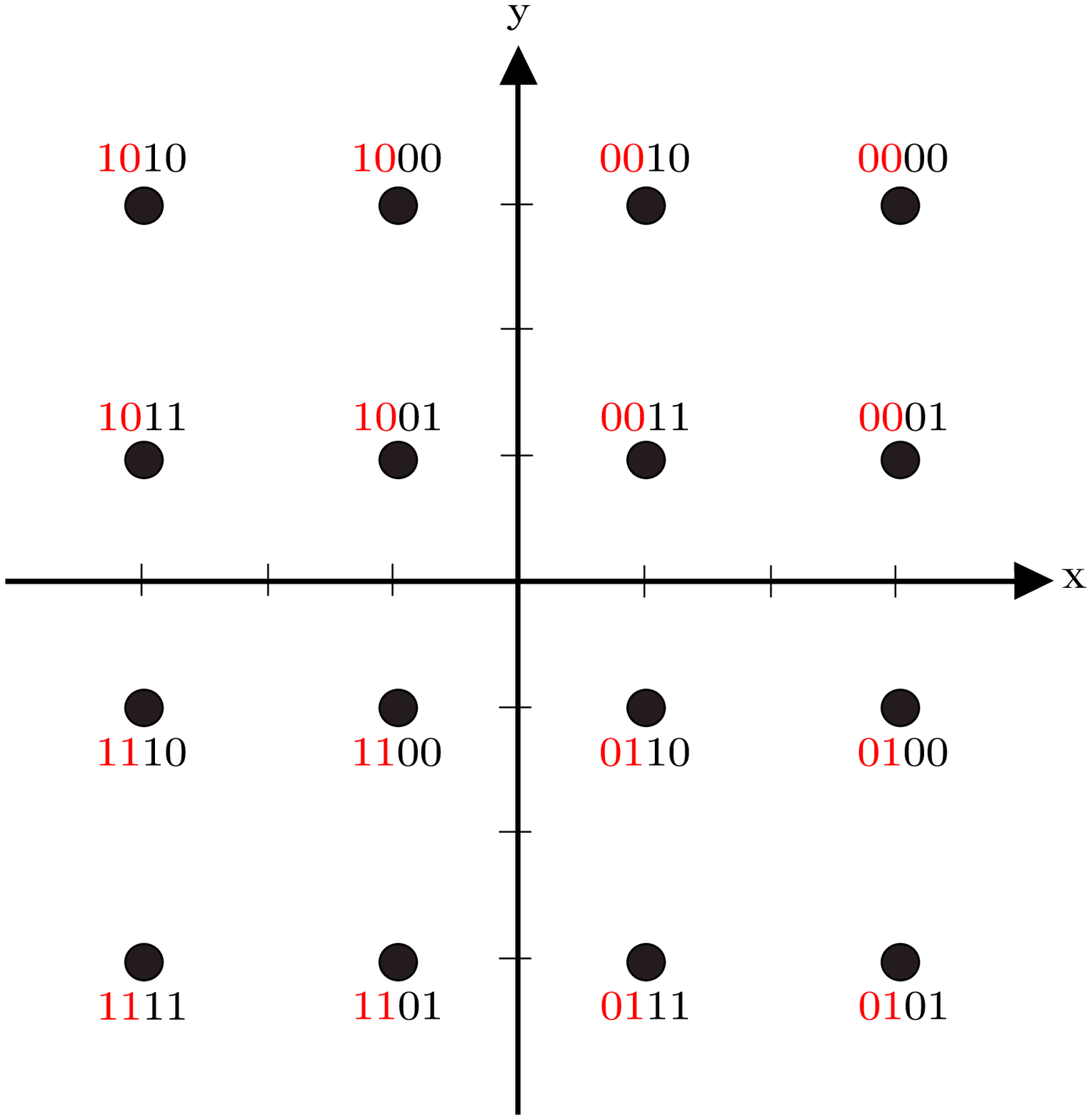}
\caption{Constellation diagram of a NOMA symbol, (4-QAM modulation for both users).}
\label{fig:qamconst}
\end{figure}

In this section, we present the expressions for the conditional BER, as an average of the probabilities of error for each bit of user $U_1$. First of all, the index representation of each bit will be $b_{1i}$, where $i$ is the index of the first user's bits, with $i \in \{1,2,...,B_1\}$ and $B_1= \log_2M_1$. If we take as an example the case of $M_1 = M_2 = 4$, we can create a constellation diagram such as in Fig. \ref{fig:qamconst}, where the $B_1 = 2$ first bits will represent $U_1$'s symbols and the rest will represent $U_2$'s ones.

It should be mentioned that the use of Gray coding allows the appearance of patterns, which is quite useful in the analysis of the multiple antenna scenario. Specifically, due to Gray coding, the BER expression is affected only by the real part of the interference and the noise for the first $B_1/2$ bits. In contrast, it is affected by the imaginary part of them for the remaining $B_1/2$ bits of the first user, as we can notice in Fig. \ref{fig:qamconst}. This phenomenon, in combination with the fact that $U_1$'s bits remain unchanged inside each quadrant, generates two properties that can be exploited for the benefit of the detection of $\hat{s}_1$ in (\ref{eq:hat_s_1}). Initially, the $x$ and $y$ axes can be used as the decision boundaries of the two groups of $U_1$'s bits, respectively. Secondarily, the computation of the distances between all the feasible symbols and $\hat{s}_1$ can be done either by choosing only four symmetrical symbols from the constellation, as seen clearly in Fig. \ref{fig:qamconst}. Due to these properties, the probability of error for the first $B_1/2$ bits is equal to the probability of error with the second $B_1/2$ bits, with the only difference being the decision boundary we consider. Consequently, in our case, the probability of error for $b_{11}$ is equal to the probability of error for $b_{12}$. Thus, the expression of BER for $U_1$ in the case of $M_1 = M_2 = 4$ is 
\begin{equation}
\begin{split}
{}&{}
P_{e}^{(1)} = \frac{1}{4}\bigg[Q(g_1 \left( 1, -1, -1 \right)) +Q(g_1 \left( 1, -1, 1 \right))\\&
\;\;\;\;\;\;\;\;\;\;\;\; + Q(g_1 \left( 1, 1, -1 \right)) + Q(g_1 \left( 1, 1, 1 \right))\bigg].
\end{split}
\label{eq:P_bi_1_4_4}
\end{equation}
with $g_1$ as defined in \eqref{eq:P_bi_1_expr_3}.

\subsection{$U_2$ EXPRESSION}

Previously, we explained the connection between the bits of the first user's symbol that are estimated during the SIC and the bits of the $U_2$ symbol. Since both users' constellations use Gray coding, we know that the probability of error for the $b_{21}$ and $b_{22}$ bits are equal. Thus, the average conditional BER expression of the second user for $M_1 = M_2 = 4$ can be written as
\begin{equation}
\begin{split}
{}&{}
P_{e}^{(2)} = \frac{1}{2}\bigg[2Q(g^+_2(0,1)) - Q(g^+_2(1,1)) + Q(g^+_2(2,1))\\&
\;\;\;\;\;\;\;\;\;\;\;+ Q(g^-_2(1,1)) - Q(g^-_2(2,1))\bigg],
\end{split}
\label{eq:P_bi_2_4_4}
\end{equation}
with $g_{2}^\pm$ as defined in \eqref{eq:P_bi_2_exprFINAL}.

\section*{ACKNOWLEDGMENT}
This work was carried out within the framework of the French collaborative project "Covera5Ge" supported by DGA and whose partners are CentraleSupélec, ENENSYS Technologies, and Siradel.

\bibliographystyle{ieeetr}
\bibliography{BERlibrary.bib}

\end{document}